\documentclass{mnras}

\usepackage[T1]{fontenc}
\usepackage{ae,aecompl,multirow}

\usepackage{graphicx}	
\usepackage{amsmath}	
\usepackage{amssymb}	
\usepackage{color}

\title[Radiative rates and opacity calculations in \ion{Ce}{ii--iv}]{Multiconfiguration Dirac-Hartree-Fock radiative parameters for emission lines in Ce II -- IV ions and cerium opacity calculations for kilonovae}

\author[H. Carvajal Gallego et al.]{
H. Carvajal Gallego,$^{1}$
P. Palmeri,$^{1}$
and P. Quinet,$^{1,2}$\thanks{E-mail: Pascal.Quinet@umons.ac.be}
\\
$^{1}$Physique Atomique et Astrophysique, Universit\'e de Mons, B-7000 Mons, Belgium\\
$^{2}$IPNAS, Universit\'e de Li\`ege, Sart Tilman, B-4000 Li\`ege, Belgium\\
}

\date{Accepted XXX. Received YYY; in original form ZZZ}

\pubyear{2018}

\begin{document}
\label{firstpage}
\pagerange{\pageref{firstpage}--\pageref{lastpage}}
\maketitle

\begin{abstract}
Large-scale calculations of atomic structures and radiative properties have been carried out for singly, doubly- and trebly ionized cerium. For this purpose, the purely relativistic multiconfiguration Dirac-Hartree-Fock (MCDHF) method was used, taking into account the effects of valence-valence and core-valence electronic correlations in detail. The results obtained were then used to calculate the expansion opacities characterizing the kilonovae observed as a result of neutron star mergers. Comparisons with previously published experimental and theoretical studies have shown that the results presented in this work are the most complete currently available, in terms of quantity and quality, concerning the atomic data and monochromatic opacities for Ce II, Ce III and Ce IV ions.
\end{abstract}




\section{Introduction}

On August 17, 2017, the LIGO/VIRGO collaboration detected, for the first time, gravitational waves produced by the coalescence of neutron stars (Abbott {\it et al} 2017). This detection was named GW170817. During such coalescence, the very hot and radioactive ejected matter is the site of nuclear reactions leading to the formation of a large amount of atomic species heavier than iron, such as lanthanides for example (Kasen {\it et al} 2017). This phenomenon, known as kilonova, makes it possible to study the origin of these heavy elements as well as their interesting properties, in particular their high opacity due to the multitude of transitions resulting from their complex atomic structures characterised by configurations with unfilled 4f orbital.

The atomic structures and radiative processes characterising the lanthanides have already been the subject of various theoretical and experimental studies in recent years. It would be too tedious to give an exhaustive list of these works here but it is worth recalling that, about 20 years ago, we undertook a systematic and detailed analysis of the spectroscopic properties of the first four ionization stages of lanthanide atoms, from neutral to trebly charged ions. For this purpose, the pseudo-relativistic Hartree-Fock (HFR) method (Cowan 1981) including core-polarization effects (HFR+CPOL), as described by Quinet {\it et al} (1999, 2002), was intensively used to compute the radiative parameters (wavelengths, transition probabilities, oscillator strengths) in many different lanthanide ions. In order to allow a wide dissemination of the new results obtained, we created the DREAM database (Database on Rare-Earths At Mons University)\footnote{https://hosting.umons.ac.be/html/agif/databases/dream.html} which currently contains spectroscopic information for more than 72000 spectral lines belonging to neutral, singly-, doubly-, and trebly-ionized lanthanide atoms. These investigations gave rise to about fifty publications, the summary of which is given in a recent review paper (Quinet \& Palmeri 2020).

Following the GW170817 detection, a collaboration between Lithuanian and Japanese researchers was set up to analyse the light emitted by the kilonova. More precisely, large-scale atomic calculations were undertaken to model the electronic structures and radiative processes characterising heavy ions, including some lanthanide ions. To do this, the purely relativistic theoretical methods MCDHF (Multiconfiguration Dirac-Hartree-Fock) (Grant 2007, Froese Fischer {\it et al} 2016) and HULLAC (Hebrew University Lawrence Livermore Atomic Code) (Bar Shalom {\it et al} 2001) were used to obtain a very large number of new fundamental parameters related to the spectral lines belonging to some lowly ionized lanthanide atoms, namely the Nd II - IV ions (Gaigalas {\it et al} 2019), the Er III ion (Gaigalas {\it et al} 2020), the ions from Pr II to Gd II (Rad{\v{z}}i{\={u}}t{\.{e}} {\it et al} 2020), and the ions from La I - IV to Lu I - IV (Z = 71) (Tanaka {\it et al} 2020). In these works, the opacities due to the ions considered were also estimated.

Unfortunately, atomic data for lanthanide elements are still too sparse, both in quantity and quality, to accurately model the kilonova emission spectra, especially with regard to the opacity and the light curve. The main objective of our work is to make a new contribution to this field of research by considering one of the most abundant lanthanide elements, namely cerium, in its first few ionization stages. More precisely, we report large-scale atomic structure calculations for Ce II, Ce III and Ce IV ions performed using the MCDHF method in which we considered the valence-valence and core-valence correlation effects in great detail. The radiative parameters obtained in these calculations were then used to evaluate the corresponding opacities in the astrophysical context of kilonovae.

\section{Atomic data calculations}

\subsection{Computational procedure}

We used the fully relativistic multiconfiguration Dirac-Hartree-Fock (MCDHF) method described by Grant (2007) and Froese Fischer {\it et al} (2016) for computing the atomic structures and radiative parameters in Ce II--IV ions, with the latest version of GRASP (General Relativistic Atomic Structure Program), i.e. GRASP2018 (Froese Fischer {\it et al} 2019). As a reminder, in this approach, the atomic state functions (ASFs), $\Psi$, are represented by a superposition of configuration state functions (CSFs), $\Phi$, with the same parity, $P$, total angular momentum, and total magnetic quantum numbers, $J$ and $M$ :

\begin{equation}
\psi(\gamma PJM)=\sum_{j=1}^{N_{CSF}}c_{j}\Phi(\gamma_{j}PJM),
\end{equation}

\noindent where the label $\gamma_{j}$ represents all the other quantum numbers needed to univoquely specify CSFs that are $jj$-coupled Slater determinants built from one-electron spin–orbitals. The configuration mixing coefficients \textit{$c_{j}$} are obtained through the diagonalisation of the Dirac-Coulomb Hamiltonian

\begin{equation}
H_{DC} = \sum_{i=1}^{N}(c \mathbf{\alpha_{i}}.\mathbf{p_{i}}+(\beta_{i}-1)c^{2}+V(r_{i}))+\sum_{i>j}^{N}\dfrac{1}{r_{ij}},
\end{equation}

\noindent where $V$($r$) is the monopole part of the electron-nucleus interaction.

Finally, the high-order relativistic effects, i.e. the Breit interaction, QED self-energy and vacuum polarization effects are
incorporated in the relativistic configuration interaction (RCI) step of the GRASP2018 package.

In the present work, different physical models were applied to each ion in order to optimize the wave functions and the corresponding energy levels by gradually increasing the basis of CSFs, and thus taking into account more correlations. In a first step, valence-valence (VV) models, in which single and double excitations (SD) of valence electrons, i.e. occupying open subshells of conﬁgurations from a multi-reference (MR) to a set of active spectroscopic orbitals were considered in order to generate the CSF expansions. These sets of active orbitals are denoted $n$s, $n$'p, $n$"d, ... where $n$, $n$', $n$", ... are the maximum principal quantum numbers considered for each azimuthal quantum number $l$. In a second step, core-valence (CV) models consisted in adding single and double excitations from core-orbitals such as 4d, 5s and 5p to the VV models. The details of calculations performed in Ce II, Ce III, and Ce IV are given below.

\subsection{Ce II}

The ground configuration of singly ionised cerium is 4f5d$^2$. In our calculations, the MR consisted in 4f5d$^2$, 4f$^2$6p, 4f5d6s, 4f$^3$ odd-parity configurations and 4f$^2$5d, 4f$^2$6s, 4f5d6p, 4f6s6p, 5d$^3$ even-parity configurations.
The orbitals were optimised in several steps. For the odd-conﬁgurations, we first optimised the spectroscopic orbitals from 1s to 5d on the ground conﬁguration. Then, we optimised 6s and 6p separately by fixing the orbitals obtained previously. For even-parity, all orbitals, from 1s to 6p, were optimised together using the whole set of configurations.

A first VV model (VV1) was built by adding to the MR configurations, single and double excitations from 5d, 4f, 6s, and 6p to the active orbitals {6s,6p,5d,5f,5g} (J = 1/2 to 15/2). Only the new orbitals, 5f and 5g, were optimised, the other ones being kept to their values obtained before. The same strategy was used to build more elaborate VV models by successively considering the following sets of active orbitals : {6s,6p,6d,6f,6g} (VV2), {7s,7p,7d,7f,7g} (VV3), {8s,8p,8d,8f,7g} (VV4), {9s,9p,9d,8f,7g} (VV5), and {10s,10p,9d,8f,7g} (VV6), always limited to J = 1/2 -- 15/2. It was verified that it was not necessary to go beyond the VV6 model with respect to the valence-valence correlation, as the results obtained showed a very good convergence, in terms of wavelengths and radiative rates, as shown in Figures 1 and 2 for a few selected transitions. Therefore a CV1 model was then built by adding SD excitations from 5s and 5p core orbitals to 4f, 5d, 6s and 6p from the VV6 model.

In Table 1, we report the comparison between the lowest energy levels calculated in this work and the corresponding experimental values taken from the NIST compilation (Kramida {\it et al} 2020). We can clearly see that the deviations generally decrease when moving from VV6 to CV1 model. This is also illustrated in Figure 3 and 4 where the distributions of relative deviations from experimental level energies are shown for the VV6 and CV1 models, respectively. As can be seen in these two figures, the discrepancies between our theoretical results and the experimental energy levels are substantially reduced when going from VV6 to CV1, the average deviations being found to be $\Delta E/E$ = 0.39 $\pm$ 0.38 and 0.18 $\pm$ 0.15, respectively. However, we notice that, in a few sporadic instances, our calculations still struggle to reproduce the experimental levels. This is particularly the case for the 4f5d($^3$F)6s $^4$F multiplet for which the CV1 calculations give energies much lower than those compiled at NIST. This is essentially due to the difficulty of theoretically modeling these levels for which a significant mixing is observed with states belonging to the 4f5d$^2$ configuration. It can be expected that an even more elaborate CV model, including single and double excitations from the 4d core orbital, could improve the agreement with the experimental data, but such a model could not be tested in our work because of computational limitations.

The wavelengths, transition probabilities and oscillator strengths were then computed for the spectral lines involving all the Ce II atomic states obtained in the CV1 model. When comparing these results with the data taken from the DREAM database (Quinet \& Palmeri 2020), we found a good overall agreement, i.e. within 5\% for the wavelengths and within 30\% for the transition rates. A comparison between the $gA$-values obtained in the present work and those listed in the DREAM database is given in Table 2 for a selected sample of intense Ce II transitions ($gA$ $>$ 10$^8$ s$^{-1}$). The ratios between our results computed in the Babushkin ($B$) and Coulomb ($C$) gauges are also reported in the same table, showing a good agreement with a mean ratio $B$/$C$ equal to 1.15 $\pm$ 0.12.

\subsection{Ce III}

The ground configuration of Ce III is 4f$^2$. In this ion, the MR was made up of 4f$^2$, 4f6p, 5d$^2$, 5d6s even-parity configurations and 4f5d, 4f6s, 5d6p odd-parity configurations. In a first step, we optimized the orbitals from 1s to 4f on the 4f$^2$ ground configuration for values of the total angular momentum $J$ between 0 and 7. The 5d, 6s and 6p orbitals were optimized using the MR even-configurations, keeping all the other orbitals fixed. The same procedure was then carried out using the odd-parity configurations.

A second step consisted in the extension of CSF basis by completing the MR expansion by adding SD excitations from 4f, 5d, 6s and 6p to the active space of 6s, 6p, 5d, 5f and 5g orbitals, giving rise to the VV1 model. Only 5f and 5g were optimised, all the other orbitals being fixed to the values obtained before. Following the same strategy of optimizing only the newly introduced orbitals, we built more elaborate VV models, namely VV2 : 6s,6p,6d,6f,6g,6h; VV3 : 7s,7p,7d,7f,7g,7h; VV4 : 8s,8p,8d,7f,7g,7h, and VV5 : 9s,9p,8d,7f,7g,7h. The convergence of results (wavelengths and transition rates) was verified when going from VV1 to VV5 model, as shown in Figures 5 and 6 for selected lines. In the case of Ce III, two CV models were considered, each of which is an extension of the VV5 model. In the first one (CV1), SD excitations from 5s and 5p core-orbitals to 6s, 6p, 5d and 4f were allowed. The numbers of CSFs obtained in this case not being too large (54809 and 48863 for even and odd parities, respectively), it was possible to consider a more elaborate CV model including excitations from the 4d subshell. More precisely, in this CV2 model, we added the 4d$^2$ $\rightarrow$ 4f$^2$ excitation, the importance of which was demonstrated by Froese Fischer and Godefroid (2019) to better reproduce the experimental energy levels of the 4f$^2$ ground configuration in Ce III.

A comparison of our energy level values obtained in VV5, CV1 and CV2 models with experimental data compiled at NIST (Kramida {\it et al} 2020) is given in Table 3. It can be seen that, as expected, our CV2 calculations are in much better agreement with experiment. This is illustrated in Figures 7-9 where the distributions of relative differences are shown, the mean deviation going from $\Delta E/E$ = 0.30 $\pm$ 0.16 to 0.29 $\pm$ 0.15 and 0.11 $\pm$ 0.09 when considering our VV5, CV1 and CV2 models, respectively.

In Table 4, we compare the radiative transition probabilities obtained using our CV2 model with those listed in the DREAM database (Quinet \& Palmeri 2020) for the 50 most intense lines ($gA$ $>$ 10$^7$ s$^{-1}$). A rather good overall agreement is observed when comparing both sets of data, the mean ratio $gA$$_{This~work}$/$gA$$_{DREAM}$ being found to be equal to 0.90 $\pm$ 0.56.

\subsection{Ce IV}

The atomic structure of Ce IV is quite simple with only one electron outside a Xe-like ionic core. In this case, our MR was made up of 4f, 5f, 6p, and 5d, 5g, 6s, 6d, 7s, 7d and 8s configurations in the odd and even parities, respectively. In order to obtain the optimised orbitals, we proceeded as we did for Ce II and Ce III, so that the orbitals from 1s to 4f were optimized on the 4f ground configuration, while the other ones were optimised one by one using all MR configurations.

Different VV models were then built by adding SD excitations from valence subshells to the following active orbitals : 8s,6p,5d,5f,5g (MR); 8s,6p,6d,6f,6g,6h (VV2); 8s,7p,7d,7f,7g,7h (VV3); 8s,8p,8d,7f,7g,7h (VV4); 9s,9p,8d,7f,7g,7h (VV5). The convergence of results (wavelengths and transition probabilities) from MR to VV5 models is shown in Figures 10 and 11 for specific cases. Therefore, we used the VV5 model as a starting point to construct a CV1 model by adding SD excitations from 5s and 5p core orbitals to 6s, 7s, 8s, 6p, 5d, 4f, 5f and 5g orbitals. This led to an improvement of the agreement between our theoretical energy levels and the experimental values taken from the NIST compilation (Kramida {\it et al} 2020), as shown in Figures 12 and 13. Indeed, when going from VV5 to CV1 model, the mean energy difference was found to be reduced from 0.078 $\pm$ 0.020 to 0.016 $\pm$ 0.020.

\section{Opacity calculations}

\subsection{Formalism of expansion opacities}

To evaluate bound-bound opacities in a rapidly expanding environment, such as in the ejecta from neutron star mergers, the formalism of expansion opacities is commonly used (Karp {\it et al} 1977; Eastman \& Pinto 1993; Kasen {\it et al} 2006). The main feature of this formalism is that the contributions of a large number of lines to the monochromatic opacity are approximated by a discretization involving the summation
of lines falling within a spectral width, while the radiative transfer is considered in Sobolev (1960) approximation.

The bound-bound opacity is thus calculated using the following expression :

\begin{equation}
\kappa^{bb}(\lambda) = {1 \over \rho c t} \sum_{l} {\lambda_{l} \over \Delta\lambda} (1 - e^{-\tau_{l}}),
\end{equation}

\noindent where $\lambda$ (in \AA) is the central wavelength within the region of width $\Delta \lambda$, $\lambda_l$ are the wavelengths of the lines appearing in this range, $\tau_l$ are the corresponding optical depths, $c$ (in cm/s) is the speed of light, $\rho$ (in g/cm$^3$) is the density of the ejected gas and $t$ (in s) is the elapsed time since ejection.

The optical depth can be expressed by (Sobolev 1960) :

\begin{equation}
\tau_{l} = {\pi e^{2} \over m_{e} c} f_{l} n_{l} t \lambda_{l},
\end{equation}

\noindent where $e$ (in C) is the elementary charge, $m_e$ (in g) is the electron mass, $f_l$ (dimensionless) is the oscillator strength, and $n_l$ (in cm$^{-3}$) is the density of the lower level of the transition.

In this formalism, the local thermodynamic equilibrium (LTE) is assumed. Indeed, although the low density of ejecta from neutron star mergers at $t$ = 1 day is not high enough for many collisions to establish LTE, in optically thick plasmas ($\tau$ $>>$ 1), the radiation field tends towards a blackbody law and the radiative transitions lead to LTE level populations. It is then possible to express $n_l$ using the Boltzmann distribution as :

\begin{equation}
n_l = {g_l \over g_0} n e^{- {E_l/k_B T}},
\end{equation}

\noindent where $k_B$ is the Boltzmann constant (in cm$^{-1}$K$^{-1}$), $T$ (in K) is the temperature, $g_l$ and $E_l$ (in cm$^{-1}$) are respectively the statistical weight and the energy of the lower level of the transition, and $g_0$ is the statistical weight of the ground level for the ion considered.

We can then write the optical depth as

\begin{equation}
\tau_l = {\pi e^2 \over m_e c} ({n \lambda_l t \over g_0}) g_l f_l e^{-E_l/k_B T}.
\end{equation}

For a charge state $j$, the ionic density $n_j$ (in cm$^{-3}$) is obtained using the Saha equation

\begin{equation}
{n_j \over n_{j-1}} = {U_j(T) U_e(T) \over U_{j-1}(T) n_e} e^{-\chi_{j-1}/k_B T},
\label{saha}
\end{equation}

\noindent where $n_{j-1}$ is the ionic density in the $j$-1 charge stage, $n_e$ is the electron density, $\chi_{j-1}$ is the ionization potential of the ion $j$-1, $U_j$($T$) and $U_{j-1}$($T$) are the partition functions for charge stages $j$ and $j$-1, respectively, computed using all the energy levels,  $E_i^{(j)}$ and $E_i^{(j-1)}$, and their statistical weights, $g_i^{(j)}$ and $g_i^{(j-1)}$, belonging to the corresponding ions :

\begin{equation}
U_j(T) = \sum_{i} g_i^{(j)} e^{-E_i^{(j)} / k_B T},
\label{partj}
\end{equation}

\begin{equation}
U_{j-1}(T) = \sum_{i} g_i^{(j-1)} e^{-E_i^{(j-1)} / k_B T}.
\label{partjm1}
\end{equation}

The electronic partition function, $U_e$, is given by :

\begin{equation}
U_e(T) = 2 ({m_e k_B T \over 2 \pi \hbar^2})^{3/2}.
\label{parte}
\end{equation}

\subsection{Opacities for Ce II--IV ions}

On the basis of the above-mentioned expressions, opacities were calculated using the most reliable atomic data obtained in the present work for Ce II, Ce III, and Ce IV ions, i.e. using the most elaborate MCDHF CV models. The temperatures used to calculate these opacities were $T$ = 5000 K (Ce II), 10000 K (Ce III) , and 15000 K (Ce IV) while the density was fixed at $\rho$ = 10$^{-13}$ g.cm$^{-3}$ and the time after merger was $t$ = 1 day. These values correspond to those generally assumed for an ejecta mass of $M_{ej}$ $\sim$ 10$^{-2}$ $M_{Sun}$ and velocity $v$ $\sim$ 0.1 $c$ (Tanaka {\it et al} 2020). Moreover, a pure Ce gas was assumed.

In Figures 14, 15 and 16, we show the expansion opacities calculated by using transition rates of Ce II, Ce III, and Ce IV, respectively. The blue lines correspond to the results deduced from the best atomic data computed in the present paper, while the red lines in Figures 14 and 15 display the results obtained using the radiative parameters listed in the DREAM database (Quinet and Palmeri 2020). When looking at Figure 14 it can be clearly observed that, for Ce II, the behaviour of opacity is similar whether we use our atomic data or those taken from DREAM, both curves having a maximum at about 4400 \AA ~(4355 \AA ~using our data and 4395 \AA ~using DREAM). However, a more detailed inspection shows that the opacities calculated in the present work are systematically slightly higher than those obtained from the DREAM data. This is not only due to the differences between the atomic data but also to the larger number of lines considered in the present  study for the opacity calculations. This is much more marked in the case of Ce III, where we see, from Figure 15, a much greater difference between the opacities computed in our work and those obtained with the DREAM data. To quantify things, the numbers of transitions considered in the different opacity calculations are compared in Table 5. For completeness, the atomic data used in our opacity calculations are given in Tables 6, 7 and 8 for Ce II, Ce III and Ce IV, respectively. These tables are available in their entirety in a machine-readable version at the Centre de Donn\'ees Astronomiques de Strassbourg (CDS) through anonymous ftp to cdsarc.u-strasbg.fr (130.79.128.5) or via http://cdsweb.u-strasbg.fr.

In order to test the influence of the discretization step $\Delta \lambda$ in the opacity calculations, we have redone the calculations with $\Delta \lambda$ = 5 \AA ~(instead of $\Delta \lambda$ = 10 \AA). The results obtained for Ce II and Ce III are shown in Figures 17 and 18, respectively. As seen from these figures, when compared to Figures 14 and 15, a smaller value of $\Delta \lambda$ improves the spectral resolution of monochromatic opacities but does not change the conclusions drawn hereabove.

In addition, we have looked at the variation of the expansion opacities with the elapsed time since ejection $t$ as shown in Figures~19 and 20 for, respectively, Ce~II and Ce~III. Three different elapsed times were considered, i.e. $t=$ 0.5, 1 and 1.5~days, fixing the density ($\rho=$~10$^{-13}$~g.cm$^{-3}$),  the temperature ($T=$5000~K  for Ce~II; $T=$10000~K for Ce~III) and the spectral region width ($\Delta \lambda =$~5~\AA). One can clearly see that the opacities globally decrease with $t$ as expected from a dilution effect. The latter is related to the hypothesis of an isotropic velocity gradient that produces a photon mean free path proportionnal to the expansion time $t$ (Karp $et~al$, 1977).

Finally, we have determined from the Saha equation (Eqs.(\ref{saha}--\ref{parte})) the different temperatures of maximum ionic density for the different ions studied, i.e. Ce~II--IV, using our MCDHF level energies. They were $\sim$3500~K for Ce~II, $\sim$6500~K for Ce~III and $\sim$12000~K for Ce~IV. These can be compared to those typically assumed by Tanaka~\textit{et al.}~(2018, 2020) for respectively all the second, third and fourth lanthanide spectra, $i.e.$ $\sim$5000~K (II) , $\sim$10000~K (III) and $\sim$15000~K (IV).

\section{Conclusion}

In our work, we have focused on the first three ions of cerium, for which we have modeled the atomic structures and calculated the radiative transition rates using the purely relativistic multiconﬁguration Dirac-Hartree-Fock (MCDHF) method. In order to obtain atomic parameters as reliable as possible, different physical models including valence-valence and core-valence correlation were considered in a systematic and progressive manner for each ion. The quality of the results obtained could be highlighted thanks to the convergence of the parameters calculated from the different theoretical models and also thanks to the good agreement observed by comparing the theoretical energy levels with the available experimental data, the deviation being generally of the order of a few percent. This allowed us to determine a new set of reliable transition probabilities and oscillator strengths for a large amount of spectral lines in Ce II, Ce III, and Ce IV, that were then used to calculate astrophysical opacities in the context of kilonovae.

More precisely, the atomic parameters obtained in the present work for 30194 electric dipole (E1) transitions in Ce II, 77044 E1 transitions in Ce III, and 37 E1 transitions in Ce IV, were used to compute expansion opacities required for radiative transfer simulations of kilonovae, radioactively powered by electromagnetic emission from neutron star mergers. Our results were compared with data deduced from transition rates previously published, such as those compiled in the DREAM database. The latter being systematically based on a much smaller number of lines and on less elaborate theoretical approaches than the one adopted in our investigation, we can conclude that the present paper constitutes a substantial and reliable contribution to the study of opacities affecting the emission spectra of kilonovae produced during neutron star mergers.

\section*{Acknowledgements}

P.P. and P.Q. are, respectively, Research Associate and Research Director of the Belgian Fund for Scientific Research F.R.S.-FNRS. Financial support from this organization is gratefully acknowledged.

\section*{Data availability}

The data (Tables 6--8) underlying this article are available at the Centre de Donn\'ees Astronomiques de Strasbourg (CDS) through anonymous ftp to cdsarc.u-strasbg.fr (130.79.128.5) or via http://cdsweb.u-strasbg.fr.




\bibliographystyle{mnras}

\def\bibindent{1em}



%
%


\clearpage

\clearpage

   \begin{table*}
   \centering
      \caption{Comparison between the energies calculated in the present work using VV6 and CV1 models and the experimental values compiled at NIST for the lowest levels of Ce II.}
         \label{KapSou}
         $$
         \begin{tabular}{cccccccccc}
            \hline
Configuration        & Term & $J$     & $E$ (EXP)$^a$     & $E$ (VV6)$^b$       & Diff (VV6)$^c$   & $E$ (CV1)$^d$    & Diff (CV1)$^e$   \\
                     &      &       & (cm$^{-1}$) & (cm$^{-1}$)   & (\%)           & (cm$^{-1}$) & (\%)           \\
\hline
4f($^{2} \!$F)5d$^{2}$($^{3} \!$F) & $^{4} \!$H & 7/2  & 0.00     & 0.00     &        &  0.00   &        \\
4f($^{2} \!$F)5d$^{2}$($^{3} \!$F) & *          & 9/2  & 987.61   & 811.00   & -17.8  & 948.87  & -3.9   \\
4f($^{2} \!$F)5d$^{2}$($^{3} \!$F) & $^{4} \!$I & 9/2  & 1410.30  & 1357.93  & -3.7   & 1541.43 & 9.3    \\
4f($^{2} \!$F)5d$^{2}$($^{3} \!$F) & *          & 7/2  & 1873.93  & 1767.98  & -5.6   & 631.84  & -66.3  \\
4f($^{2} \!$F)5d$^{2}$($^{3} \!$F) & *          & 1/2  & 2140.47  & 4987.67  & 133.1  & 2397.24 & 12.0   \\
4f5d($^{1} \!$G)6s                 & *          & 9/2  & 2382.24  & 2179.86  & -8.5   & 2513.26 & 5.5    \\
4f($^{2} \!$F)5d$^{2}$($^{3} \!$F) & $^{4} \!$I & 11/2 & 2563.23  & 2460.44  & -4.0   & 2690.38 & 5.0    \\
4f($^{2} \!$F)5d$^{2}$($^{3} \!$F) & $^{4} \!$H & 9/2  & 2581.25  & 2146.16  & -16.8  & 2170.24 & -15.9  \\
4f5d($^{3} \!$F)6s                 & $^{4} \!$F & 3/2  & 2595.64  & 2604.74  & 0.4    & 818.16  & -68.5  \\
4f5d($^{3} \!$F)6s                 & *          & 5/2  & 2634.66  & 2728.54  & 3.6    & 2212.84 & -16.0  \\
4f5d($^{1} \!$G)6s                 & *          & 7/2  & 2641.55  & 2307.06  & -12.6  & 1802.34 & -31.8  \\
4f($^{2} \!$F)5d$^{2}$($^{3} \!$F) & $^{4} \!$H & 11/2 & 2879.70  & 2128.11  & -26.1  & 2464.12 & -14.4  \\
4f5d($^{3} \!$F)6s                 & $^{4} \!$F & 5/2  & 3363.43  & 3355.00  & -0.2   & 1140.16 & -66.1  \\
4f($^{2} \!$F)5d$^{2}$($^{3} \!$F) & $^{2} \!$S & 1/2  & 3508.47  & 6485.63  & 84.9   & 4081.20 & 16.3   \\
4f$^{2}$($^{3} \!$H)6s             & *          & 9/2  & 3593.88  & 3141.68  & -12.6  & 3123.56 & -13.1  \\
4f($^{2} \!$F)5d$^{2}$($^{3} \!$F) & *          & 7/2  & 3703.59  & 4226.40  & 14.1   & 3414.08 & -7.8   \\
4f($^{2} \!$F)5d$^{2}$($^{3} \!$F) & $^{4} \!$D & 3/2  & 3745.48  & 3788.32  & 1.1    & 3494.10 & -6.7   \\
4f($^{2} \!$F)5d$^{2}$($^{3} \!$F) & $^{4} \!$I & 13/2 & 3793.63  & 3715.10  & -2.1   & 4011.29 & 5.7    \\
4f$^{2}$($^{3} \!$H)6s             & $^{4} \!$H & 7/2  & 3854.01  & 11208.63 & 190.8  & 5634.18 & 46.2   \\
4f5d($^{3} \!$H)6s                 & $^{4} \!$H & 7/2  & 3995.46  & 3890.09  & -2.6   & 2326.19 & -41.8  \\
4f$^{2}$($^{3} \!$H)6s             & $^{4} \!$H & 9/2  & 4165.55  & 11635.24 & 179.4  & 5928.87 & 42.3   \\
4f($^{2} \!$F)5d$^{2}$($^{3} \!$F) & $^{4} \!$F & 3/2  & 4201.89  & 4991.97  & 18.8   & 4709.7  & 12.1   \\
4f($^{2} \!$F)5d$^{2}$($^{3} \!$F) & $^{4} \!$H & 13/2 & 4203.93  & 3165.58  & -24.7  & 3574.05 & -15.0  \\
4f5d($^{3} \!$G)6s                 & *          & 7/2  & 4266.39  & 11490.49 & 169.3  & 4236.6  & -0.7   \\
4f5d($^{3} \!$G)6s                 & *          & 5/2  & 4322.70  & 6931.16  & 60.4   & 5839.46 & 35.1   \\
4f5d($^{3} \!$F)6s                 & $^{4} \!$F & 7/2  & 4459.87  & 4854.53  & 8.9    & 2207.12 & -50.5  \\
4f5d($^{3} \!$G)6s                 & $^{4} \!$G & 5/2  & 4511.26  & 4875.11  & 8.1    & 2885.69 & -36.0  \\
4f5d($^{3} \!$H)6s                 & $^{4} \!$H & 9/2  & 4523.03  & 4588.87  & 1.5    & 3544.1  & -21.6  \\
4f($^{2} \!$F)5d$^{2}$($^{3} \!$F) & *          & 5/2  & 4737.37  & 4370.4   & -7.7   & 4362.7  & -7.9   \\
4f($^{2} \!$F)5d$^{2}$($^{3} \!$F) & *          & 3/2  & 4844.64  & 7824.72  & 61.5   & 3702.73 & -23.6  \\
4f5d($^{3} \!$H)6s                 & *          & 11/2 & 4910.96  & 9220.44  & 87.8   & 4502.16 & -8.3   \\
4f($^{2} \!$F)5d$^{2}$($^{3} \!$F) & $^{4} \!$F & 5/2  & 5010.87  & 5676.31  & 13.3   & 5250.87 & 4.8    \\
4f5d($^{1} \!$D)6s                 & *          & 5/2  & 5118.80  & 6373.62  & 24.5   & 3592.03 & -29.8  \\
4f($^{2} \!$F)5d$^{2}$($^{3} \!$F) & *          & 1/2  & 5283.03  & 3057.42  & -42.1  & 5051.83 & -4.4   \\
4f($^{2} \!$F)5d$^{2}$($^{3} \!$F) & $^{4} \!$G & 7/2  & 5437.42  & 6826.51  & 25.5   & 6263.4  & 15.2   \\
4f($^{2} \!$F)5d$^{2}$($^{3} \!$F) & $^{4} \!$I & 15/2 & 5455.85  & 4992.28  & -8.5   & 5329.56 & -2.3   \\
4f$^{2}$($^{3} \!$H)6s             & $^{4} \!$H & 11/2 & 5513.70  & 12798.22 & 132.1  & 7135.59 & 29.4   \\
4f$^{2}$($^{3} \!$H)6s             & $^{2} \!$H & 9/2  & 5616.74  & 12882.35 & 129.4  & 7187.84 & 28.0   \\
4f5d($^{3} \!$H)6s                 & $^{4} \!$H & 11/2 & 5651.36  & 5556.79  & -1.7   & 3607.82 & -36.2  \\
4f5d($^{3} \!$F)6s                 & $^{4} \!$F & 5/2  & 5675.76  & 5166.75  & -9.0   & 4388.47 & -22.7  \\
4f($^{2} \!$F)5d$^{2}$($^{3} \!$F) & *          & 7/2  & 5716.22  & 5891.49  & 3.1    & 3214.14 & -43.8  \\
4f5d($^{3} \!$G)6s                 & *          & 9/2  & 5819.11  & 6154.25  & 5.8    & 6794.95 & 16.8   \\
    \hline
\end{tabular}
$$
\begin{list}{}{}
\item[$^{\mathrm{a}}$] Experimental energy levels from the NIST database (Kramida {\it et al.} 2020)
\item[$^{\mathrm{b}}$] Calculated MCDHF energy levels obtained in the present work using the VV6 model
\item[$^{\mathrm{c}}$] Percentage relative difference : 100 $\times$ (E(VV6) - E(EXP)) / E(EXP)
\item[$^{\mathrm{d}}$] Calculated MCDHF energy levels obtained in the present work using the CV1 model
\item[$^{\mathrm{e}}$] Percentage relative difference : 100 $\times$ (E(CV1) - E(EXP)) / E(EXP)
\end{list}
\end{table*}

\clearpage

   \begin{table*}
   \centering
      \caption{Comparison between the transition probabilities ($gA$) calcuated in the present work and those listed in the DREAM database for the most intense Ce II lines.}
         \label{KapSou}
         $$
         \begin{tabular}{cccccccccc}
            \hline
Wavelength$^a$    & \multicolumn{3}{c}{Lower level$^b$} & \multicolumn{3}{c}{Upper level$^b$} & $gA$ (DREAM)$^c$ & $gA$ (This work)$^d$ & $B$/$C$ (This work)$^e$ \\
(\AA)         & $E$ (cm$^{-1}$) & $P$ & $J$ & $E$ (cm$^{-1}$) & $P$ & $J$ & (s$^{-1}$) & (s$^{-1}$) &         \\
\hline
2154.159      &  5651 & (o) & 11/2 & 52059 & (e) &  9/2 & 1.19 $\times$ 10$^8$ & 2.51 $\times$ 10$^8$ & 1.37 \\
3201.710      &  6913 & (o) & 13/2 & 38138 & (e) & 11/2 & 2.35 $\times$ 10$^9$ & 2.79 $\times$ 10$^9$ & 1.26 \\
3234.888      &  7234 & (o) & 11/2 & 38138 & (e) & 11/2 & 7.92 $\times$ 10$^8$ & 9.68 $\times$ 10$^8$ & 1.24 \\
3272.251      &  5651 & (o) & 11/2 & 36203 & (e) &  9/2 & 8.72 $\times$ 10$^8$ & 7.17 $\times$ 10$^8$ & 1.25 \\
3274.867      &  5676 & (o) &  9/2 & 36203 & (e) &  9/2 & 5.70 $\times$ 10$^8$ & 2.19 $\times$ 10$^8$ & 1.27 \\
3304.821      & 11742 & (o) & 11/2 & 41992 & (e) & 13/2 & 6.90 $\times$ 10$^8$ & 5.93 $\times$ 10$^8$ & 1.40 \\
3488.549      &  7059 & (o) &  9/2 & 35716 & (e) & 11/2 & 2.96 $\times$ 10$^8$ & 4.33 $\times$ 10$^8$ & 1.32 \\
3513.845      & 11388 & (o) &  7/2 & 39838 & (e) &  5/2 & 1.68 $\times$ 10$^8$ & 1.68 $\times$ 10$^8$ & 1.60 \\
3517.377      &  7294 & (o) & 13/2 & 35716 & (e) & 11/2 & 8.73 $\times$ 10$^8$ & 2.92 $\times$ 10$^8$ & 0.97 \\
3526.682      & 12326 & (o) & 13/2 & 40674 & (e) & 11/2 & 6.70 $\times$ 10$^8$ & 6.56 $\times$ 10$^8$ & 1.04 \\
3534.045      &  4204 & (o) & 13/2 & 32492 & (e) & 11/2 & 4.45 $\times$ 10$^8$ & 1.69 $\times$ 10$^8$ & 1.04 \\
3560.802      &  5456 & (o) & 13/2 & 33531 & (e) & 13/2 & 1.05 $\times$ 10$^9$ & 7.84 $\times$ 10$^8$ & 1.08 \\
3577.456      &  3794 & (o) & 13/2 & 31738 & (e) & 11/2 & 8.49 $\times$ 10$^8$ & 3.21 $\times$ 10$^8$ & 1.05 \\
3623.737      & 14404 & (o) & 15/2 & 41992 & (e) & 13/2 & 1.08 $\times$ 10$^9$ & 8.77 $\times$ 10$^8$ & 1.04 \\
3653.664      &  3794 & (o) & 13/2 & 31156 & (e) & 13/2 & 4.35 $\times$ 10$^8$ & 9.30 $\times$ 10$^8$ & 1.15 \\
3655.844      &  2563 & (o) & 11/2 & 29909 & (e) &  9/2 & 4.71 $\times$ 10$^8$ & 5.20 $\times$ 10$^8$ & 1.07 \\
3709.287      &  4204 & (o) & 13/2 & 31156 & (e) & 13/2 & 5.34 $\times$ 10$^8$ & 1.22 $\times$ 10$^8$ & 1.12 \\
3728.018      &  5676 & (o) &  9/2 & 32492 & (e) & 11/2 & 2.05 $\times$ 10$^8$ & 9.08 $\times$ 10$^7$ & 1.14 \\
3728.417      &  5456 & (o) & 15/2 & 32269 & (e) & 15/2 & 7.56 $\times$ 10$^8$ & 6.16 $\times$ 10$^8$ & 1.19 \\
3801.526      &  7234 & (o) & 11/2 & 33531 & (e) & 13/2 & 2.37 $\times$ 10$^9$ & 2.36 $\times$ 10$^9$ & 1.21 \\
3803.096      &  2880 & (o) & 11/2 & 29167 & (e) &  9/2 & 2.44 $\times$ 10$^8$ & 1.46 $\times$ 10$^8$ & 1.12 \\
3811.596      & 11742 & (o) & 11/2 & 37971 & (e) & 11/2 & 7.90 $\times$ 10$^8$ & 9.50 $\times$ 10$^8$ & 1.15 \\
3848.592      &  4204 & (o) & 13/2 & 30180 & (e) & 13/2 & 4.83 $\times$ 10$^8$ & 5.95 $\times$ 10$^8$ & 1.19 \\
3889.982      &  5456 & (o) & 15/2 & 31156 & (e) & 13/2 & 6.26 $\times$ 10$^8$ & 5.31 $\times$ 10$^8$ & 1.08 \\
3903.929      & 12366 & (o) &  9/2 & 37974 & (e) &  7/2 & 4.58 $\times$ 10$^8$ & 2.41 $\times$ 10$^8$ & 1.32 \\
3908.404      &  6913 & (o) & 13/2 & 32492 & (e) & 11/2 & 6.01 $\times$ 10$^8$ & 6.15 $\times$ 10$^8$ & 1.09 \\
3919.803      &  5651 & (o) & 11/2 & 31156 & (e) & 13/2 & 3.07 $\times$ 10$^8$ & 4.05 $\times$ 10$^8$ & 1.24 \\
3933.730      &  5676 & (o) &  9/2 & 31090 & (e) & 11/2 & 1.54 $\times$ 10$^9$ & 1.16 $\times$ 10$^9$ & 1.24 \\
3938.084      &  4523 & (o) &  9/2 & 29909 & (e) &  9/2 & 1.29 $\times$ 10$^8$ & 1.06 $\times$ 10$^8$ & 1.18 \\
3942.745      &  6913 & (o) & 13/2 & 32269 & (e) & 15/2 & 2.32 $\times$ 10$^9$ & 2.54 $\times$ 10$^9$ & 1.28 \\
4003.767      &  7523 & (o) & 11/2 & 32492 & (e) & 11/2 & 8.19 $\times$ 10$^8$ & 3.84 $\times$ 10$^8$ & 1.04 \\
4019.057      &  8176 & (o) &  5/2 & 33050 & (e) &  3/2 & 1.22 $\times$ 10$^8$ & 2.00 $\times$ 10$^8$ & 1.02 \\
4027.690      & 17171 & (o) & 11/2 & 41992 & (e) & 13/2 & 1.47 $\times$ 10$^9$ & 7.59 $\times$ 10$^8$ & 1.10 \\
4036.108      &  8281 & (o) &  5/2 & 33050 & (e) &  3/2 & 1.02 $\times$ 10$^8$ & 3.71 $\times$ 10$^7$ & 1.08 \\
4075.700      &  5651 & (o) & 11/2 & 30180 & (e) & 13/2 & 8.68 $\times$ 10$^8$ & 1.26 $\times$ 10$^9$ & 1.27 \\
4083.222      &  5651 & (o) & 11/2 & 30135 & (e) & 11/2 & 7.47 $\times$ 10$^8$ & 3.31 $\times$ 10$^8$ & 1.13 \\
4083.629      & 16192 & (o) &  9/2 & 40674 & (e) & 11/2 & 1.17 $\times$ 10$^9$ & 1.20 $\times$ 10$^9$ & 1.15 \\
4123.869      &  6913 & (o) & 13/2 & 31156 & (e) & 13/2 & 1.03 $\times$ 10$^9$ & 1.05 $\times$ 10$^9$ & 1.17 \\
4142.825      &  5676 & (o) &  9/2 & 29807 & (e) &  7/2 & 1.26 $\times$ 10$^8$ & 2.58 $\times$ 10$^8$ & 1.04 \\
4175.233      &  5965 & (o) &  7/2 & 29909 & (e) &  9/2 & 1.08 $\times$ 10$^8$ & 2.08 $\times$ 10$^8$ & 1.12 \\
4190.618      &  7234 & (o) & 11/2 & 31090 & (e) & 11/2 & 2.21 $\times$ 10$^8$ & 1.45 $\times$ 10$^9$ & 1.15 \\
4193.065      &  5965 & (o) &  7/2 & 29807 & (e) &  7/2 & 1.68 $\times$ 10$^8$ & 1.10 $\times$ 10$^8$ & 1.07 \\
4251.595      & 19947 & (o) &  3/2 & 43461 & (e) &  5/2 & 1.08 $\times$ 10$^8$ & 1.00 $\times$ 10$^8$ & 1.19 \\
4255.781      &  5676 & (o) &  9/2 & 29167 & (e) &  9/2 & 3.20 $\times$ 10$^8$ & 3.55 $\times$ 10$^8$ & 1.18 \\
4286.920      &  9139 & (o) &  3/2 & 42459 & (e) &  3/2 & 3.41 $\times$ 10$^8$ & 1.94 $\times$ 10$^8$ & 1.13 \\
4302.654      & 10641 & (o) &  5/2 & 33876 & (e) &  3/2 & 1.66 $\times$ 10$^8$ & 3.67 $\times$ 10$^8$ & 1.11 \\
4390.274      & 11742 & (o) & 11/2 & 34513 & (e) & 13/2 & 5.06 $\times$ 10$^8$ & 1.72 $\times$ 10$^8$ & 1.00 \\
4882.463      & 12326 & (o) & 13/2 & 32802 & (e) & 11/2 & 5.49 $\times$ 10$^8$ & 1.98 $\times$ 10$^8$ & 0.94 \\
4971.494      & 14404 & (o) & 15/2 & 34513 & (e) & 13/2 & 5.90 $\times$ 10$^8$ & 3.69 $\times$ 10$^8$ & 0.94 \\
\hline
\end{tabular}
$$
\begin{list}{}{}
\item[$^{\mathrm{a}}$] Wavelengths deduced from experimental energy levels.
\item[$^{\mathrm{b}}$] Experimental energy levels from the NIST database (Kramida {\it et al.} 2020). The values rounded to the unit are given. (e) and (o) stand for even and odd parity, respectively.
\item[$^{\mathrm{c}}$] HFR+CPOL values taken from the DREAM database (Quinet and Palmeri 2020)
\item[$^{\mathrm{d}}$] MCDHF values obtained in the present work using the CV1 model (Babushkin gauge)
\item[$^{\mathrm{e}}$] Ratio between the MCDHF $gA$-values obtained in the Babushkin and Coulomb gauges
\end{list}
\end{table*}

\clearpage

   \begin{table*}
   \centering
      \caption{Comparison between the energies calculated in the present work using VV5, CV1 and CV2 models and the experimental values compiled at NIST for the lowest levels of Ce III.}
         \label{KapSou}
         $$
         \begin{tabular}{cccccccccc}
            \hline
Configuration & Term & $J$ & $E$ (EXP)$^a$   & $E$ (VV5)$^b$   & Diff (VV5)$^c$ & $E$ (CV1)$^d$ & Diff (CV1)$^e$ & $E$ (CV2)$^f$  & Diff (CV2)$^g$ \\
              &      &   & (cm$^{-1}$) & (cm$^{-1}$) & (\%)       & (cm$^{-1}$) & (\%)     & (cm$^{-1}$) & (\%)      \\
\hline
4f$^2$    & $^{3} \!$H    & 4 & 0.00     & 0         &       & 0          &          & 0             &       \\
          &               & 5 & 1528.32  & 1378.75   & -9.8  & 1405.54    & -8.0     & 1421.22       & -7.0  \\
          &               & 6 & 3127.10  & 2838.1    & -9.2  & 2893.1     & -7.5     & 2913.01       & -6.8  \\
4f5d      & $^{1} \!$G    & 4 & 3276.66  & 3796.07   & 15.8  & 6142.23    & 87.4     & 3059.3        & -6.6  \\
4f$^2$    & $^{3} \!$F    & 2 & 3762.75  & 4620.57   & 22.8  & 4357.37    & 15.8     & 4045.64       & 7.5   \\
4f5d      & $^{3} \!$F    & 2 & 3821.53  & 4758.57   & 24.5  & 6677.2     & 74.7     & 3672.46       & -3.9  \\
4f$^2$    & $^{3} \!$F    & 3 & 4764.76  & 5489.21   & 15.2  & 5245.38    & 10.1     & 4942.42       & 3.7   \\
          &               & 4 & 5006.06  & 7213.8    & 44.1  & 7124.54    & 42.3     & 4843.59       & -3.2  \\
4f5d      & $^{3} \!$H    & 4 & 5127.27  & 6101.44   & 19.0  & 8422.15    & 64.3     & 5354.48       & 4.4   \\
4f5d      & $^{3} \!$F    & 3 & 5502.37  & 6051.11   & 10.0  & 8209.92    & 49.2     & 5183.37       & -5.8  \\
4f5d      & $^{3} \!$G    & 3 & 6265.21  & 7619.49   & 21.6  & 8894.97    & 42.0     & 6101.77       & -2.6  \\
4f5d      & $^{3} \!$H    & 5 & 6361.27  & 7462.71   & 17.3  & 9786.18    & 53.8     & 6725.54       & 5.7   \\
4f5d      & $^{1} \!$D    & 2 & 6571.36  & 9118.88   & 38.8  & 9722.83    & 48.0     & 6878.92       & 4.7   \\
4f$^2$    & $^{1} \!$G    & 4 & 7120.00  & 5256.24   & -26.2 & 5196.51    & -27.0    & 6784.18       & -4.7  \\
4f5d      & $^{3} \!$F    & 4 & 7150.05  & 7703.09   & 7.7   & 9859       & 37.9     & 6848.09       & -4.2  \\
4f5d      & $^{3} \!$G    & 4 & 7836.72  & 9155.78   & 16.8  & 10444.01   & 33.3     & 7652.2        & -2.3  \\
4f5d      & $^{3} \!$H    & 6 & 8349.99  & 9455.06   & 13.2  & 11762.27   & 40.9     & 8715.52       & 4.4   \\
4f5d      & $^{3} \!$D    & 1 & 8922.05  & 11056.02  & 23.9  & 11966.75   & 34.1     & 9262.71       & 3.8   \\
4f5d      & $^{3} \!$G    & 5 & 9325.51  & 10655.51  & 14.3  & 11916.2    & 27.8     & 9157.99       & -1.8  \\
4f5d      & $^{3} \!$D    & 2 & 9900.49  & 12080.53  & 22.0  & 12923.48   & 30.5     & 10226.26      & 3.3   \\
          &               & 3 & 10126.53 & 13706.71  & 35.3  & 13058.84   & 29.0     & 10349.87      & 2.2  \\
4f5d      & $^{3} \!$P    & 0 & 11577.16 & 14495.01  & 25.2  & 15101.58   & 30.4     & 12432.59      & 7.4   \\
          &               & 1 & 11612.67 & 14515.08  & -25.0 & 15142.13   & 30.4     & 12482.14      & -7.5  \\
4f5d      & $^{1} \!$F    & 3 & 12500.72 & 11723.08  & -6.2  & 15287.13   & 22.3     & 12558.83      & 0.5   \\
4f5d      & $^{3} \!$P    & 2 & 12641.55 & 15650.83  & 23.8  & 16161.31   & 27.8     & 13514.99      & 6.9   \\
4f$^2$    & $^{1} \!$D    & 2 & 12835.09 & 15750.58  & 22.7  & 15017.94   & 17.0     & 14291.61      & 11.3  \\
4f$^2$    & $^{3} \!$P    & 0 & 16072.04 & 20380.94  & 26.8  & 19154.14   & 19.2     & 18348.75      & 14.2  \\
4f5d      & $^{1} \!$H    & 5 & 16152.32 & 18892.69  & 17.0  & 20047.74   & 24.1     & 17633.68      & 9.2   \\
4f$^2$    & $^{3} \!$P    & 1 & 16523.66 & 20778.29  & 25.7  & 19530.28   & 18.2     & 18743.92      & 13.4  \\
          &               & 2 & 17317.49 & 21432.29  & 23.8  & 20186.55   & 16.6     & 19408.47      & 12.1  \\
4f$^2$    & $^{1} \!$I    & 6 & 17420.60 & 20849.88  & 19.7  & 19890.85   & 14.2     & 20264.47      & 16.3  \\
4f5d      & $^{1} \!$P    & 1 & 18443.63 & 20265.37  & 9.9   & 21917.19   & 18.8     & 19395.68      & 5.2  \\
4f6s      & *             & 3 & 19464.46 & 26572.8   & 36.5  & 23744.59   & 22.0     & 21141.18      & 8.6  \\
4f6s      & *             & 4 & 21476.46 & 26048.22  & 21.3  & 23390.64   & 8.9      & 20763.4       & -3.3  \\
4f6s      & *             & 3 & 21849.47 & 24052.81  & 10.1  & 21361.46   & -2.2     & 18715.77      & -14.3  \\
4f$^2$    & $^{1} \!$S    & 0 & 32838.62 & 38841.31  & 18.3  & 37874.75   & 15.3     & 35380.96      & 7.7   \\
5d$^2$    & $^{3} \!$F    & 2 & 40440.20 & 64744.31  & 60.1  & 57338.98   & 41.8     & 51979.9       & 28.5  \\
          &               & 3 & 41938.54 & 66338.83  & 58.2  & 58945.61   & 40.5     & 53590.88      & 27.8  \\
          &               & 4 & 43517.46 & 68064.53  & 56.4  & 60641.11   & 39.3     & 55282.03      & 27.0  \\
5d$^2$    & $^{1} \!$D    & 2 & 46889.79 & 72296.58  & 54.2  & 63487.45   & 35.4     & 58343.95      & 24.4  \\
5d$^2$    & $^{3} \!$P    & 0 & 48075.96 & 73862.92  & 53.6  & 64021.66   & 33.2     & 59122.81      & 23.0  \\
\hline
\end{tabular}
$$
\begin{list}{}{}
\item[$^{\mathrm{a}}$] Experimental energy levels from the NIST database (Kramida {\it et al.} 2020)
\item[$^{\mathrm{b}}$] Calculated MCDHF energy levels obtained in the present work using the VV5 model
\item[$^{\mathrm{c}}$] Percentage relative difference : 100 $\times$ (E(VV5) - E(EXP)) / E(EXP)
\item[$^{\mathrm{d}}$] Calculated MCDHF energy levels obtained in the present work using the CV1 model
\item[$^{\mathrm{e}}$] Percentage relative difference : 100 $\times$ (E(CV1) - E(EXP)) / E(EXP)
\item[$^{\mathrm{f}}$] Calculated MCDHF energy levels obtained in the present work using the CV2 model
\item[$^{\mathrm{g}}$] Percentage relative difference : 100 $\times$ (E(CV2) - E(EXP)) / E(EXP)
\end{list}
\end{table*}

\clearpage

   \begin{table*}
   \centering
      \caption{Comparison between the transition probabilities ($gA$) calcuated in the present work and those listed in the DREAM database for the most intense Ce III lines.}
         \label{KapSou}
         $$
         \begin{tabular}{cccccccccc}
            \hline
Wavelength$^a$    & \multicolumn{3}{c}{Lower level$^b$} & \multicolumn{3}{c}{Upper level$^b$} & $gA$ (DREAM)$^c$ & $gA$ (This work)$^d$ & $B$/$C$ (This work)$^e$ \\
(\AA)         & $E$ (cm$^{-1}$) & $P$ & $J$ & $E$ (cm$^{-1}$) & $P$ & $J$ & (s$^{-1}$) & (s$^{-1}$) &         \\
\hline
1050          &  7120 & (e) &   4  &102369 & (o) &   3  & 1.34 $\times$ 10$^8$ & 8.20 $\times$ 10$^7$ & 0.88 \\
1092          &  4765 & (e) &   3  & 96376 & (o) &   2  & 8.01 $\times$ 10$^7$ & 2.12 $\times$ 10$^7$ & 0.68 \\
1096          &  4765 & (e) &   3  & 96022 & (o) &   3  & 5.36 $\times$ 10$^7$ & 1.43 $\times$ 10$^7$ & 0.74 \\
1102          &  3763 & (e) &   2  & 94509 & (o) &   1  & 4.77 $\times$ 10$^7$ & 1.32 $\times$ 10$^7$ & 0.66 \\
1184          & 16524 & (e) &   1  &100968 & (o) &   2  & 4.24 $\times$ 10$^7$ & 1.64 $\times$ 10$^7$ & 0.76 \\
1195          & 17317 & (e) &   2  &100968 & (o) &   2  & 1.99 $\times$ 10$^8$ & 8.17 $\times$ 10$^7$ & 0.77 \\
1205          & 12835 & (e) &   2  & 95827 & (o) &   2  & 1.23 $\times$ 10$^8$ & 5.49 $\times$ 10$^7$ & 0.81 \\
1252          & 16524 & (e) &   1  & 96376 & (o) &   2  & 3.97 $\times$ 10$^7$ & 1.89 $\times$ 10$^7$ & 0.73 \\
1726          & 12501 & (o) &   3  & 70433 & (e) &   2  & 1.36 $\times$ 10$^8$ & 1.38 $\times$ 10$^8$ & 0.77 \\
1788          & 40440 & (e) &   2  & 96376 & (o) &   2  & 9.41 $\times$ 10$^7$ & 5.60 $\times$ 10$^7$ & 0.52 \\
1799          & 40440 & (e) &   2  & 96022 & (o) &   3  & 1.42 $\times$ 10$^8$ & 1.41 $\times$ 10$^8$ & 0.74 \\
1805          & 40440 & (e) &   2  & 95827 & (o) &   2  & 2.37 $\times$ 10$^8$ & 2.85 $\times$ 10$^8$ & 0.67 \\
1837          & 43517 & (e) &   4  & 97964 & (o) &   3  & 3.64 $\times$ 10$^9$ & 4.06 $\times$ 10$^9$ & 0.65 \\
1848          &  9900 & (o) &   2  & 64011 & (e) &   2  & 5.55 $\times$ 10$^7$ & 1.98 $\times$ 10$^7$ & 0.74 \\
1849          & 41939 & (e) &   3  & 96022 & (o) &   3  & 1.71 $\times$ 10$^9$ & 1.56 $\times$ 10$^9$ & 0.68 \\
1850          & 40440 & (e) &   2  & 94509 & (o) &   1  & 1.40 $\times$ 10$^9$ & 1.72 $\times$ 10$^9$ & 0.65 \\
1856          & 10127 & (o) &   3  & 64011 & (e) &   2  & 1.46 $\times$ 10$^7$ & 1.24 $\times$ 10$^7$ & 0.82 \\
1885          & 12501 & (o) &   3  & 65551 & (e) &   3  & 3.18 $\times$ 10$^7$ & 1.57 $\times$ 10$^7$ & 0.80 \\
1890          & 12642 & (o) &   2  & 65551 & (e) &   3  & 6.92 $\times$ 10$^7$ & 2.15 $\times$ 10$^7$ & 0.55 \\
2028          & 18444 & (o) &   1  & 67730 & (e) &   0  & 2.24 $\times$ 10$^8$ & 2.56 $\times$ 10$^8$ & 0.85 \\
2030          & 50044 & (e) &   2  & 99288 & (o) &   1  & 1.81 $\times$ 10$^8$ & 1.77 $\times$ 10$^8$ & 0.65 \\
2330          &  7150 & (o) &   4  & 50058 & (e) &   4  & 1.37 $\times$ 10$^7$ & 1.04 $\times$ 10$^7$ & 1.05 \\
2480          &  6571 & (o) &   2  & 46890 & (e) &   2  & 8.86 $\times$ 10$^7$ & 1.33 $\times$ 10$^8$ & 0.73 \\
2484          &  3277 & (o) &   4  & 43517 & (e) &   4  & 3.02 $\times$ 10$^7$ & 1.05 $\times$ 10$^7$ & 0.94 \\
2490          &  9900 & (o) &   2  & 50044 & (e) &   2  & 3.78 $\times$ 10$^7$ & 4.68 $\times$ 10$^7$ & 0.83 \\
2504          & 10127 & (o) &   3  & 50058 & (e) &   4  & 6.03 $\times$ 10$^7$ & 2.20 $\times$ 10$^7$ & 0.51 \\
2504          & 10127 & (o) &   3  & 50044 & (e) &   2  & 6.61 $\times$ 10$^7$ & 6.09 $\times$ 10$^7$ & 0.79 \\
2515          &  8922 & (o) &   1  & 48674 & (e) &   1  & 3.67 $\times$ 10$^7$ & 4.42 $\times$ 10$^7$ & 0.83 \\
2553          &  8922 & (o) &   1  & 48076 & (e) &   0  & 6.43 $\times$ 10$^7$ & 6.83 $\times$ 10$^7$ & 0.80 \\
2578          &  9900 & (o) &   2  & 48674 & (e) &   1  & 1.31 $\times$ 10$^8$ & 1.45 $\times$ 10$^8$ & 0.81 \\
2630          &  5502 & (o) &   3  & 43517 & (e) &   4  & 1.36 $\times$ 10$^7$ & 1.91 $\times$ 10$^7$ & 0.61 \\
2663          & 12501 & (o) &   3  & 50044 & (e) &   2  & 1.57 $\times$ 10$^8$ & 2.06 $\times$ 10$^8$ & 0.87 \\
2673          & 12642 & (o) &   2  & 50044 & (e) &   2  & 1.02 $\times$ 10$^8$ & 1.16 $\times$ 10$^8$ & 0.69 \\
2695          & 11577 & (o) &   0  & 48674 & (e) &   1  & 2.82 $\times$ 10$^7$ & 4.73 $\times$ 10$^7$ & 0.64 \\
2697          & 11613 & (o) &   1  & 48674 & (e) &   1  & 2.70 $\times$ 10$^7$ & 2.75 $\times$ 10$^7$ & 0.69 \\
2705          & 64011 & (e) &   2  &100968 & (o) &   2  & 4.75 $\times$ 10$^8$ & 3.69 $\times$ 10$^8$ & 0.86 \\
2715          & 65551 & (e) &   3  &102369 & (o) &   3  & 2.33 $\times$ 10$^8$ & 1.13 $\times$ 10$^8$ & 0.76 \\
2719          & 10127 & (o) &   3  & 46890 & (e) &   2  & 1.52 $\times$ 10$^8$ & 1.98 $\times$ 10$^8$ & 0.85 \\
2730          &  3822 & (o) &   2  & 40440 & (e) &   2  & 8.98 $\times$ 10$^7$ & 2.21 $\times$ 10$^7$ & 0.89 \\
2744          &  5502 & (o) &   3  & 41939 & (e) &   3  & 1.10 $\times$ 10$^8$ & 8.49 $\times$ 10$^7$ & 0.59 \\
2774          & 12642 & (o) &   2  & 48674 & (e) &   1  & 2.45 $\times$ 10$^7$ & 3.16 $\times$ 10$^7$ & 0.66 \\
2802          &  7837 & (o) &   4  & 43517 & (e) &   4  & 3.16 $\times$ 10$^7$ & 4.45 $\times$ 10$^7$ & 0.81 \\
2823          & 65551 & (e) &   3  &100968 & (o) &   2  & 1.09 $\times$ 10$^9$ & 8.45 $\times$ 10$^8$ & 0.91 \\
2834          & 11613 & (o) &   1  & 46890 & (e) &   2  & 1.89 $\times$ 10$^7$ & 1.49 $\times$ 10$^7$ & 0.66 \\
2874          &  7150 & (o) &   4  & 41939 & (e) &   3  & 3.75 $\times$ 10$^7$ & 1.76 $\times$ 10$^7$ & 0.93 \\
2907          & 12501 & (o) &   3  & 46890 & (e) &   2  & 6.96 $\times$ 10$^7$ & 5.14 $\times$ 10$^7$ & 0.93 \\
2924          &  9326 & (o) &   5  & 43517 & (e) &   4  & 2.18 $\times$ 10$^8$ & 2.43 $\times$ 10$^8$ & 0.59 \\
2925          &  6265 & (o) &   3  & 40440 & (e) &   2  & 1.14 $\times$ 10$^8$ & 4.29 $\times$ 10$^8$ & 0.74 \\
3130          & 70433 & (e) &   2  &102369 & (o) &   3  & 1.45 $\times$ 10$^9$ & 1.48 $\times$ 10$^9$ & 0.67 \\
3274          & 70433 & (e) &   2  &100968 & (o) &   2  & 1.08 $\times$ 10$^8$ & 9.66 $\times$ 10$^7$ & 0.77 \\
\hline
\end{tabular}
         $$
\begin{list}{}{}
\item[$^{\mathrm{a}}$] Wavelengths deduced from experimental energy levels.
\item[$^{\mathrm{b}}$] Experimental energy levels from the NIST database (Kramida {\it et al.} 2020). The values rounded to the unit are given. (e) and (o) stand for even and odd parity, respectively.
\item[$^{\mathrm{c}}$] HFR+CPOL values taken from the DREAM database (Quinet and Palmeri 2020)
\item[$^{\mathrm{d}}$] MCDHF values obtained in the present work using the CV2 model (Babushkin gauge)
\item[$^{\mathrm{e}}$] Ratio between the MCDHF $gA$-values obtained in the Babushkin and Coulomb gauges
\end{list}
\end{table*}

\clearpage

   \begin{table*}
   \centering
      \caption{Numbers of transitions considered in different opacity calculations for Ce II, Ce III and Ce IV ions.}
         \label{KapSou}
         $$
         \begin{tabular}{cccccccccc}
            \hline
Ion           & DREAM$^a$ & Tanaka {\it et al} (2020)$^b$ & This work$^c$ \\
\hline
Ce II         & 15989     & 21239                         & 30194 \\
Ce III        & 2935      & 5556                          & 77044 \\
Ce IV         & -         & 16                            & 37 \\
\hline
\end{tabular}
         $$
\begin{list}{}{}
\item[$^{\mathrm{a}}$] HFR+CPOL calculations.
\item[$^{\mathrm{b}}$] HULLAC calculations.
\item[$^{\mathrm{c}}$] MCDHF calculations.
\end{list}
\end{table*}

\clearpage

   \begin{table*}
   \centering
      \caption{Theoretical wavelengths, oscillator strengths and transition probabilities obtained in the present work and used in opacity calculations for Ce II (The full table is available online at the Centre de Donn\'ees Astronomiques de Strasbourg (CDS)).}
         \label{KapSou}
         $$
         \begin{tabular}{cccccccccc}
            \hline
$\lambda_{vac}$~(\AA) & $E_{lo}$~(cm$^{-1}$)$^a$ & $P_{lo}$$^a$ & $J_{lo}$$^a$ & $E_{up}$~(cm$^{-1}$)$^b$ & $P_{up}$$^b$ & $J_{up}$$^b$ & log~$gf$ & $gA$~(s$^{-1}$) & $B/C$$^c$ \\
\hline
      1402.879&   818&(o)&1.5& 72100&(e)&0.5& -3.46&1.18E+06&2.24E-01\\
      1434.658&  2397&(o)&0.5& 72100&(e)&0.5& -2.79&5.25E+06&7.56E+00\\
      1457.598&  3494&(o)&1.5& 72100&(e)&0.5& -2.35&1.41E+07&6.90E+00\\
      1462.052&  3703&(o)&1.5& 72100&(e)&0.5& -4.68&6.55E+04&4.20E+00\\
      1470.177&  4081&(o)&0.5& 72100&(e)&0.5& -2.30&1.55E+07&6.75E+00\\
      ...     & ...  & ... & ... & ... & ... & ... & ... & ... & ... \\
\hline
\end{tabular}
         $$
\begin{list}{}{}
\item[$^{\mathrm{a}}$] $E_{lo}$, $P_{lo}$ and $J_{lo}$ represent the energy, the parity and the total angular momentum quantum number of the lower level.
\item[$^{\mathrm{b}}$] $E_{up}$, $P_{up}$ and $J_{up}$ represent the energy, the parity and the total angular momentum quantum number of the upper level.
\item[$^{\mathrm{c}}$] Ratio between the transition rates obtained in the Babushkin and the Coulomb gauges.
\end{list}
\end{table*}

   \begin{table*}
   \centering
      \caption{Theoretical wavelengths, oscillator strengths and transition probabilities obtained in the present work and used in opacity calculations for Ce III (The full table is available online at the Centre de Donn\'ees Astronomiques de Strasbourg (CDS)).}
         \label{KapSou}
         $$
         \begin{tabular}{cccccccccc}
            \hline
$\lambda_{vac}$~(\AA) & $E_{lo}$~(cm$^{-1}$)$^a$ & $P_{lo}$$^a$ & $J_{lo}$$^a$ & $E_{up}$~(cm$^{-1}$)$^b$ & $P_{up}$$^b$ & $J_{up}$$^b$ & log~$gf$ & $gA$~(s$^{-1}$) & $B/C$$^c$ \\\hline
       442.222&  3059&(o)&4.0&229190&(e)&4.0&-2.77&5.81E+07&1.15E+00\\
       442.376&  9263&(o)&1.0&235315&(e)&0.0&-2.53&1.01E+08&3.01E+00\\
       442.764&  3059&(o)&4.0&228913&(e)&4.0&-1.35&1.52E+09&1.28E+00\\
       443.296&  6726&(o)&5.0&232309&(e)&6.0&-2.67&7.34E+07&7.53E-01\\
       443.416&  3059&(o)&4.0&228581&(e)&4.0&-1.94&3.92E+08&1.04E+00\\
      ...     & ...  & ... & ... & ... & ... & ... & ... & ... & ... \\
\hline
\end{tabular}
         $$
\begin{list}{}{}
\item[$^{\mathrm{a}}$] $E_{lo}$, $P_{lo}$ and $J_{lo}$ represent the energy, the parity and the total angular momentum quantum number of the lower level.
\item[$^{\mathrm{b}}$] $E_{up}$, $P_{up}$ and $J_{up}$ represent the energy, the parity and the total angular momentum quantum number of the upper level.
\item[$^{\mathrm{c}}$] Ratio between the transition rates obtained in the Babushkin and the Coulomb gauges.
\end{list}
\end{table*}

   \begin{table*}
   \centering
      \caption{Theoretical wavelengths, oscillator strengths and transition probabilities obtained in the present work and used in opacity calculations for Ce IV (The full table is available online at the Centre de Donn\'ees Astronomiques de Strasbourg (CDS)).}
         \label{KapSou}
         $$
         \begin{tabular}{cccccccccc}
            \hline
$\lambda_{vac}$~(\AA) & $E_{lo}$~(cm$^{-1}$)$^a$ & $P_{lo}$$^a$ & $J_{lo}$$^a$ & $E_{up}$~(cm$^{-1}$)$^b$ & $P_{up}$$^b$ & $J_{up}$$^b$ & log~$gf$ & $gA$~(s$^{-1}$) & $B/C$$^c$ \\\hline
       501.570&     0&(o)&2.5&199374&(e)&1.5&-3.53&7.78E+06&2.84E+00\\
       503.314&     0&(o)&2.5&198683&(e)&2.5&-1.97&2.81E+08&1.67E+00\\
       504.671&     0&(o)&2.5&198149&(e)&1.5&-2.92&3.15E+07&1.55E+00\\
       508.613&  2070&(o)&3.5&198683&(e)&2.5&-4.74&4.64E+05&1.05E+00\\
       516.380&     0&(o)&2.5&193656&(e)&3.5&-2.08&2.07E+08&2.08E+00\\
      ...     & ...  & ... & ... & ... & ... & ... & ... & ... & ... \\
\hline
\end{tabular}
         $$
\begin{list}{}{}
\item[$^{\mathrm{a}}$] $E_{lo}$, $P_{lo}$ and $J_{lo}$ represent the energy, the parity and the total angular momentum quantum number of the lower level.
\item[$^{\mathrm{b}}$] $E_{up}$, $P_{up}$ and $J_{up}$ represent the energy, the parity and the total angular momentum quantum number of the upper level.
\item[$^{\mathrm{c}}$] Ratio between the transition rates obtained in the Babushkin and the Coulomb gauges.
\end{list}
\end{table*}

\clearpage

   \begin{figure*}
   \includegraphics[width=15cm,clip]{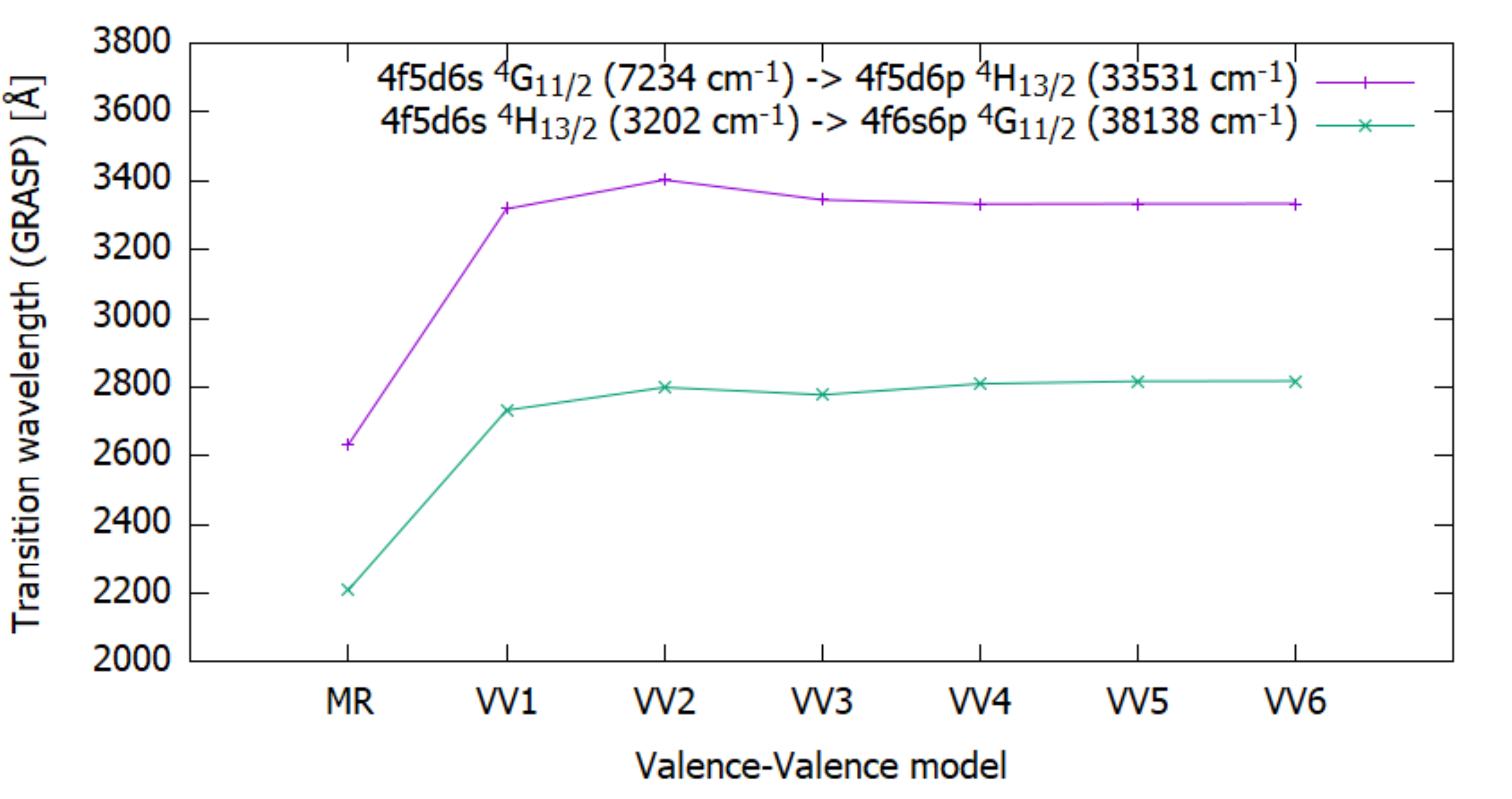}
      \caption{Convergence of wavelengths calculated in the present work using different valence-valence MCDHF models for two selected Ce II lines.}
         \label{Fig1}
   \end{figure*}

      \begin{figure*}
   \includegraphics[width=15cm,clip]{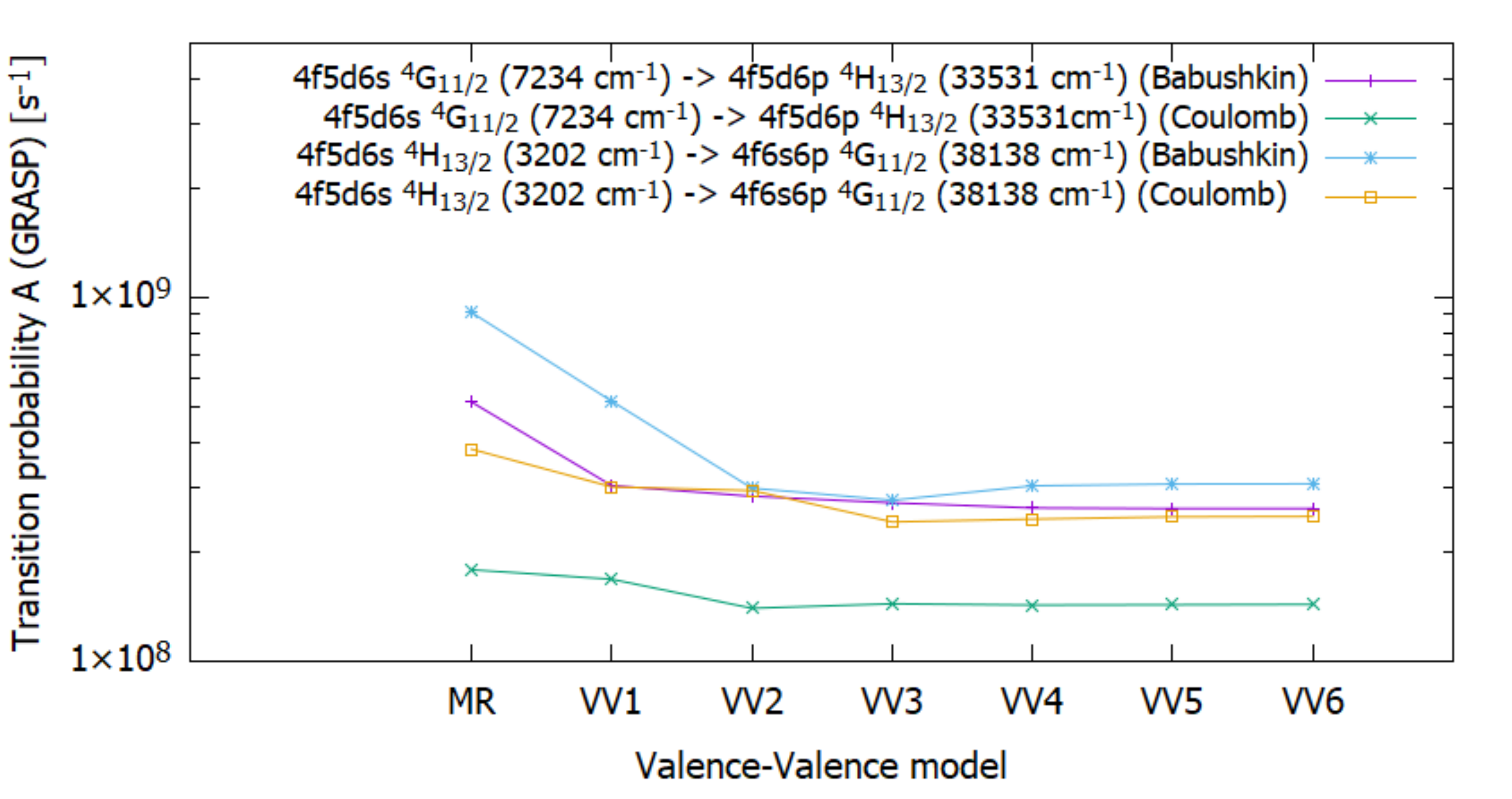}
      \caption{Convergence of transition probabilities calculated in the present work using different valence-valence MCDHF models for two selected Ce II lines.}
         \label{Fig1}
   \end{figure*}

   \begin{figure*}
   \includegraphics[width=15cm,clip]{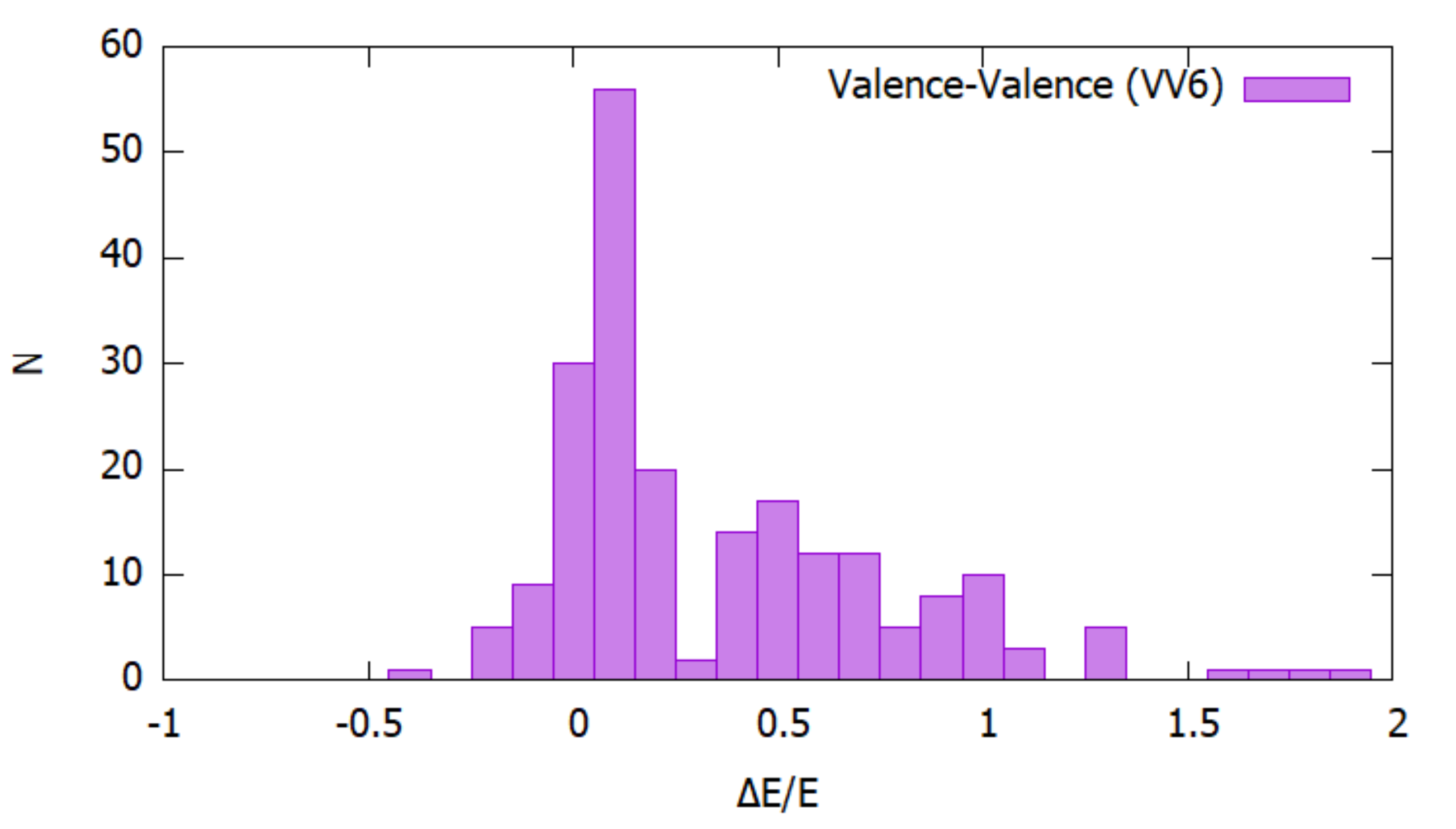}
      \caption{Distribution of energy levels ($N$) according to the mean deviation $\Delta$E$/$E with the NIST data for Ce II using our VV6 model.}
         \label{Fig1}
   \end{figure*}

      \begin{figure*}
   \includegraphics[width=15cm,clip]{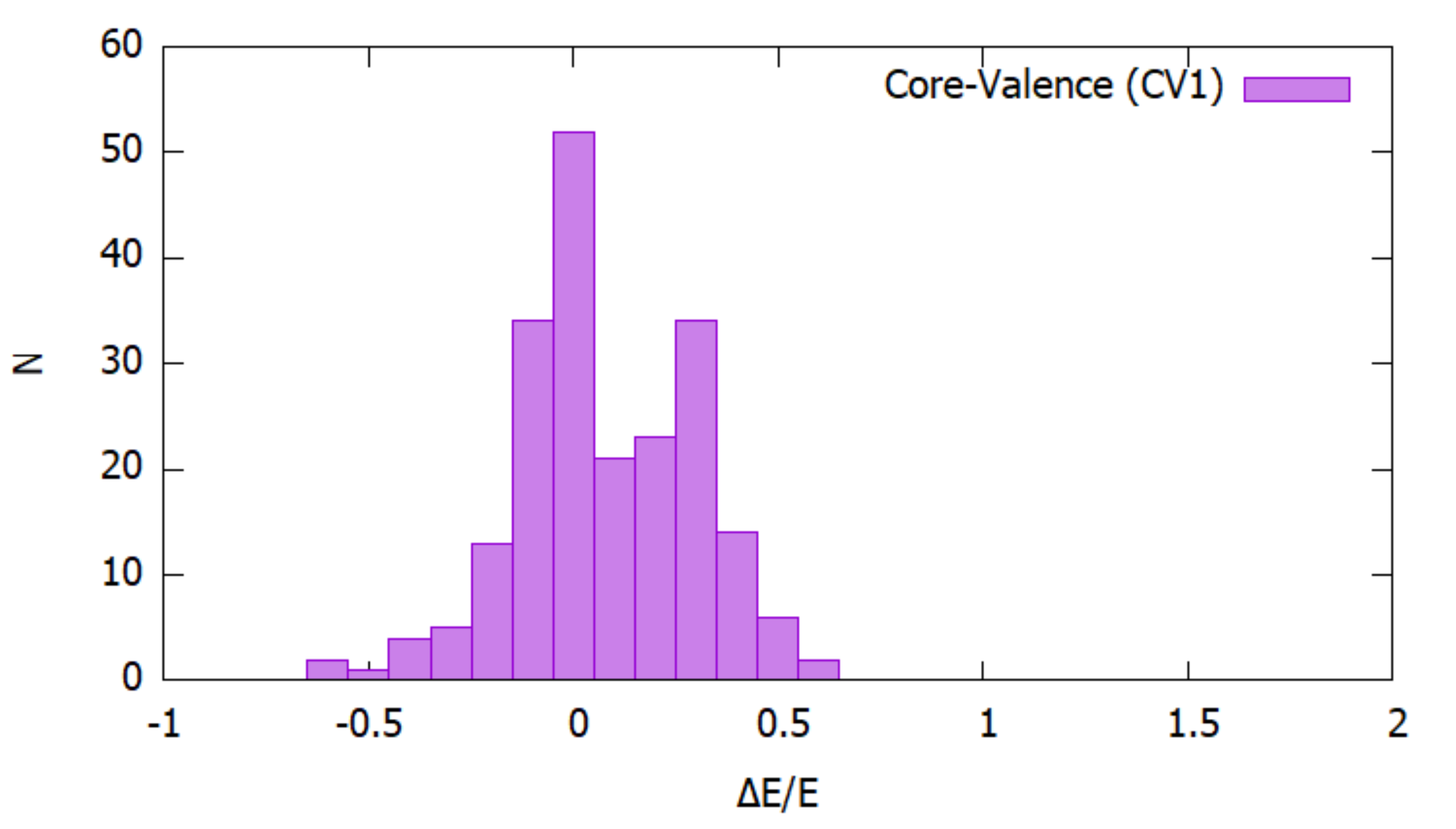}
      \caption{Distribution of energy levels ($N$) according to the mean deviation $\Delta$E$/$E with the NIST data for Ce II using our CV1 model.}
         \label{Fig1}
   \end{figure*}

     \begin{figure*}
   \includegraphics[width=15cm,clip]{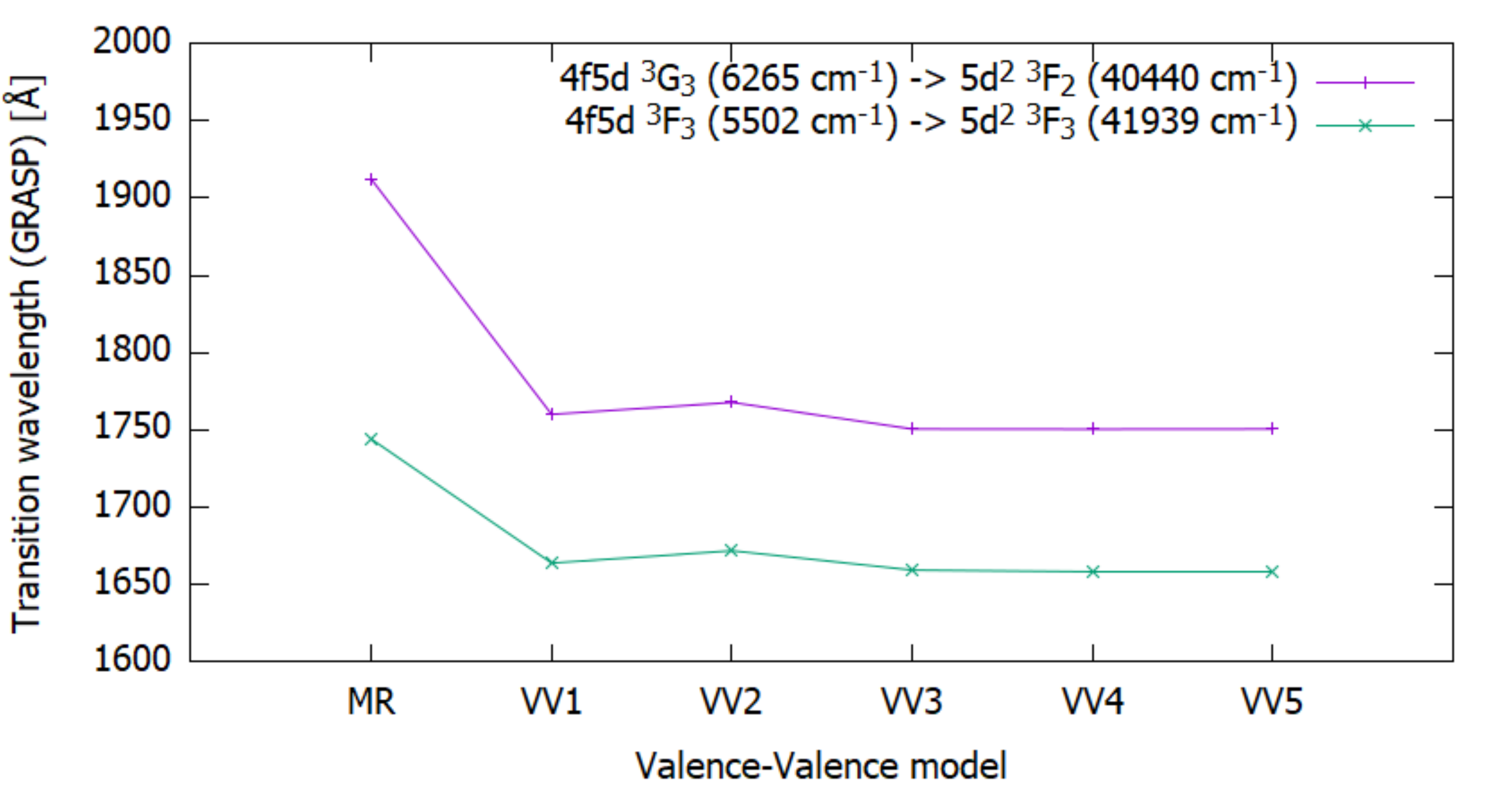}
      \caption{Convergence of wavelengths calculated in the present work using different valence-valence MCDHF models for two selected Ce III lines.}
         \label{Fig1}
   \end{figure*}

      \begin{figure*}
   \includegraphics[width=15cm,clip]{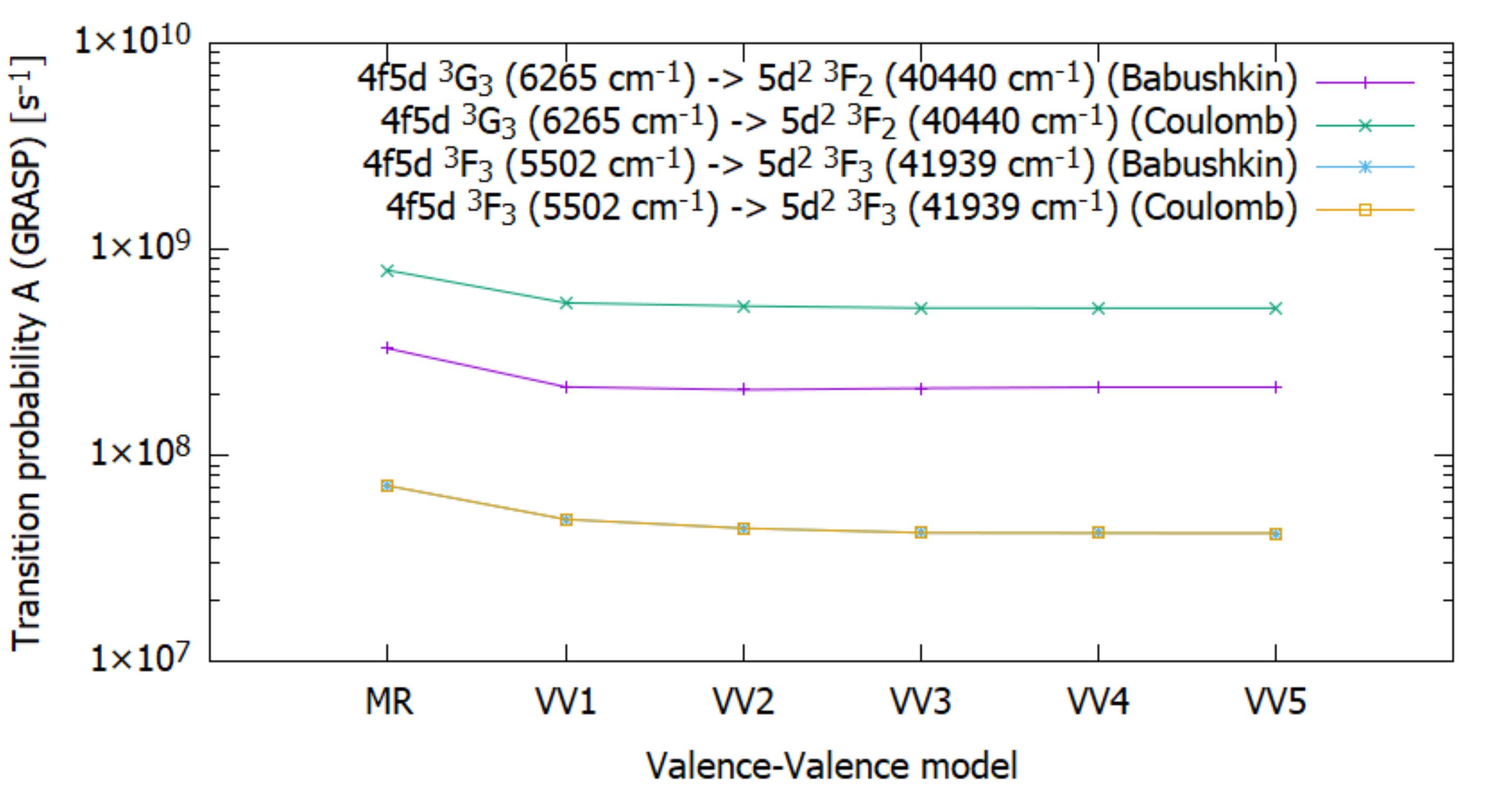}
      \caption{Convergence of transition probabilities calculated in the present work using different valence-valence MCDHF models for two selected Ce III lines.}
         \label{Fig1}
   \end{figure*}

      \begin{figure*}
   \includegraphics[width=15cm,clip]{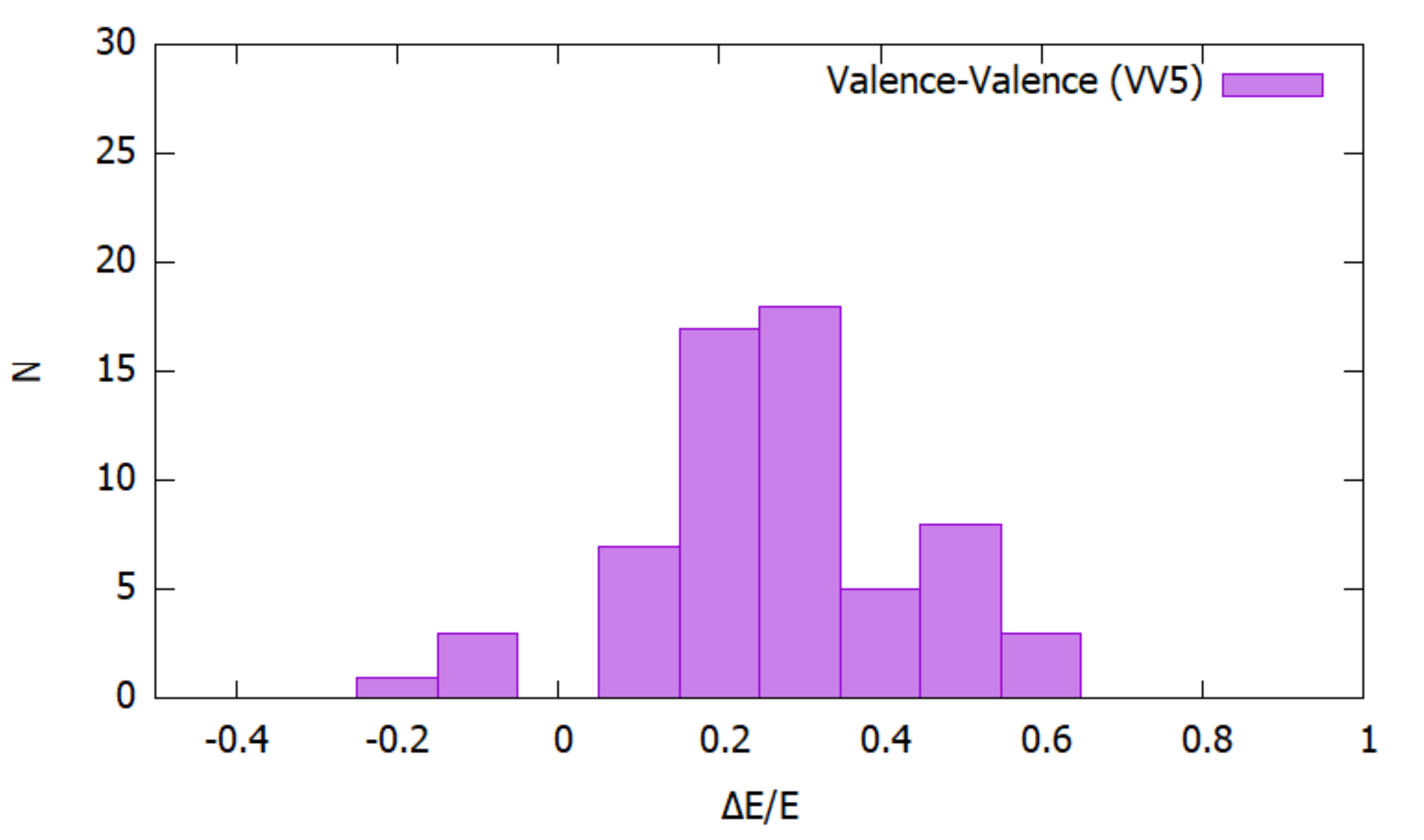}
      \caption{Distribution of energy levels ($N$) according to the mean deviation $\Delta$E$/$E with the NIST data for Ce III using our VV5 model.}
         \label{Fig1}
   \end{figure*}

      \begin{figure*}
   \includegraphics[width=15cm,clip]{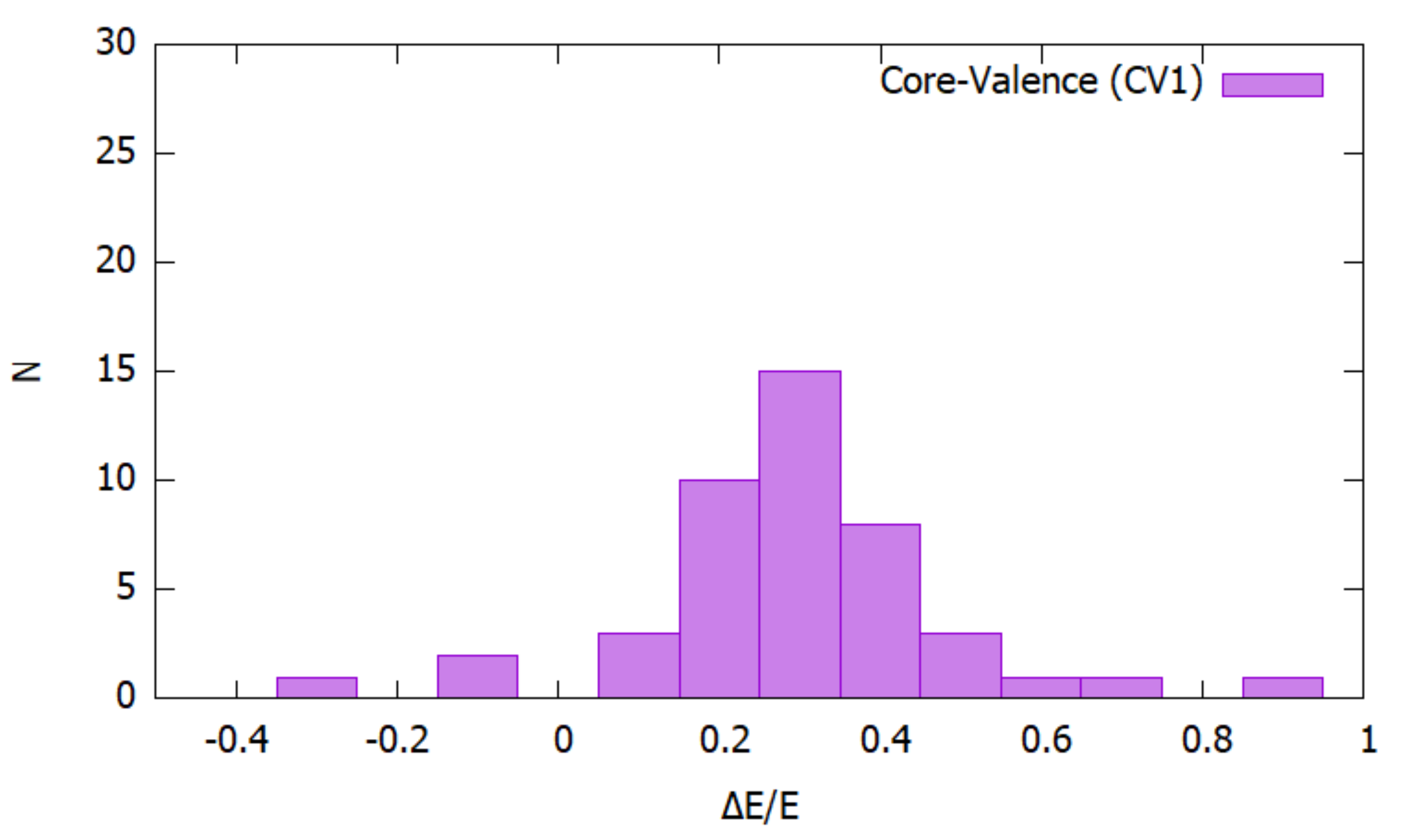}
      \caption{Distribution of energy levels ($N$) according to the mean deviation $\Delta$E$/$E with the NIST data for Ce III using our CV1 model.}
         \label{Fig1}
   \end{figure*}

         \begin{figure*}
   \includegraphics[width=15cm,clip]{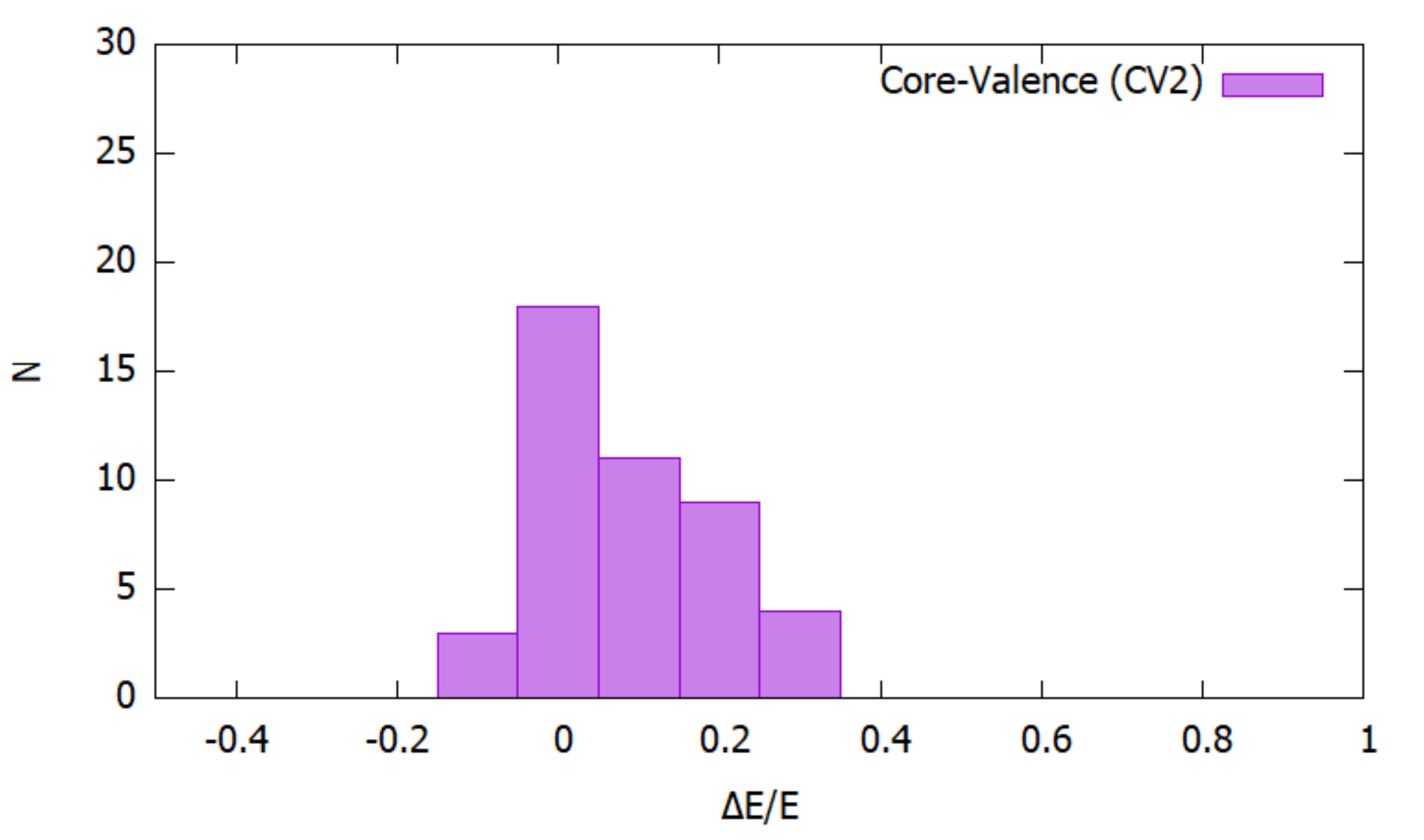}
      \caption{Distribution of energy levels ($N$) according to the mean deviation $\Delta$E$/$E with the NIST data for Ce II using our CV2 model.}
         \label{Fig1}
   \end{figure*}

      \begin{figure*}
   \includegraphics[width=15cm,clip]{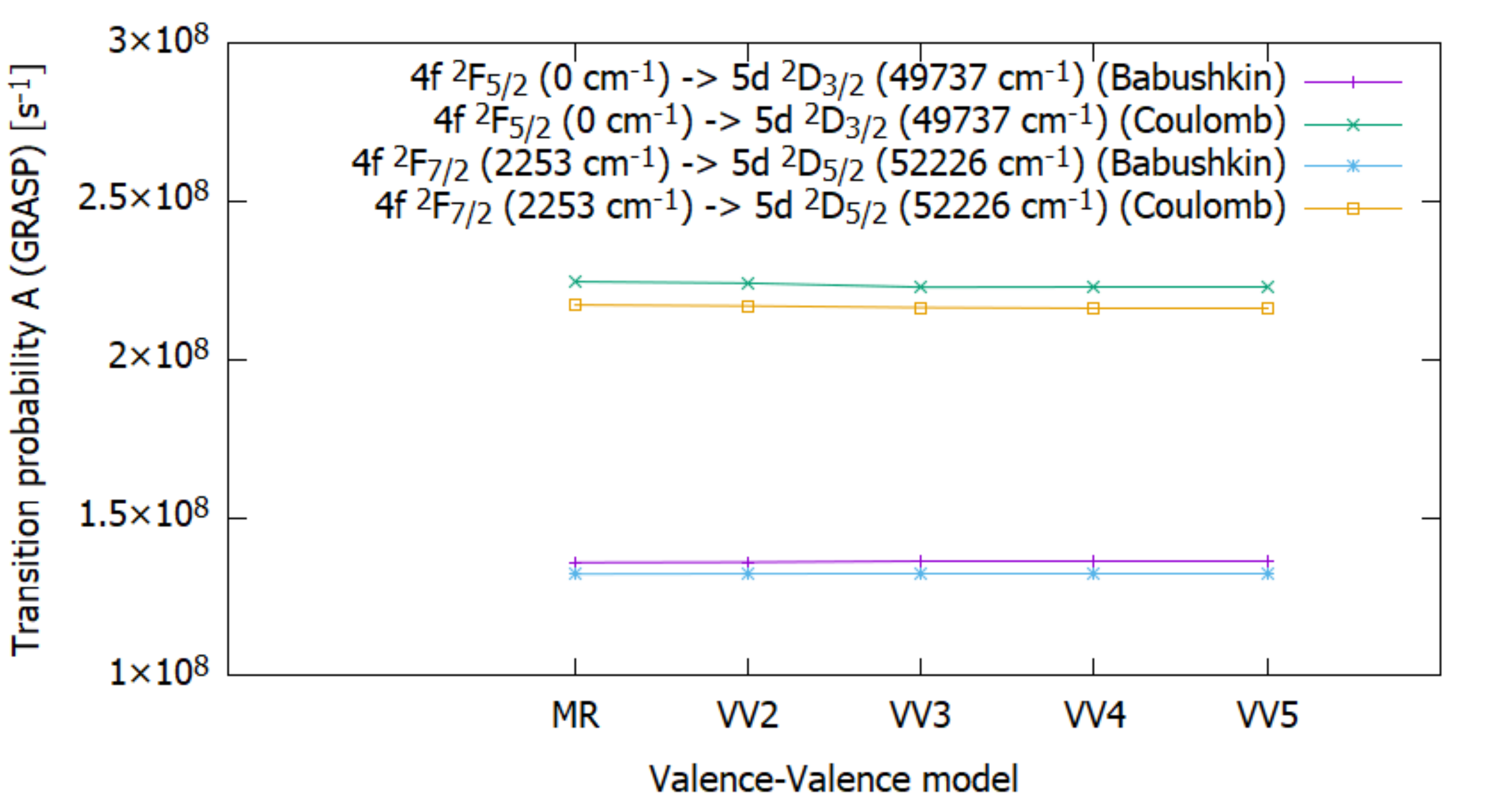}
      \caption{Convergence of wavelengths calculated in the present work using different valence-valence MCDHF models for two selected Ce IV lines.}
         \label{Fig1}
   \end{figure*}

      \begin{figure*}
   \includegraphics[width=15cm,clip]{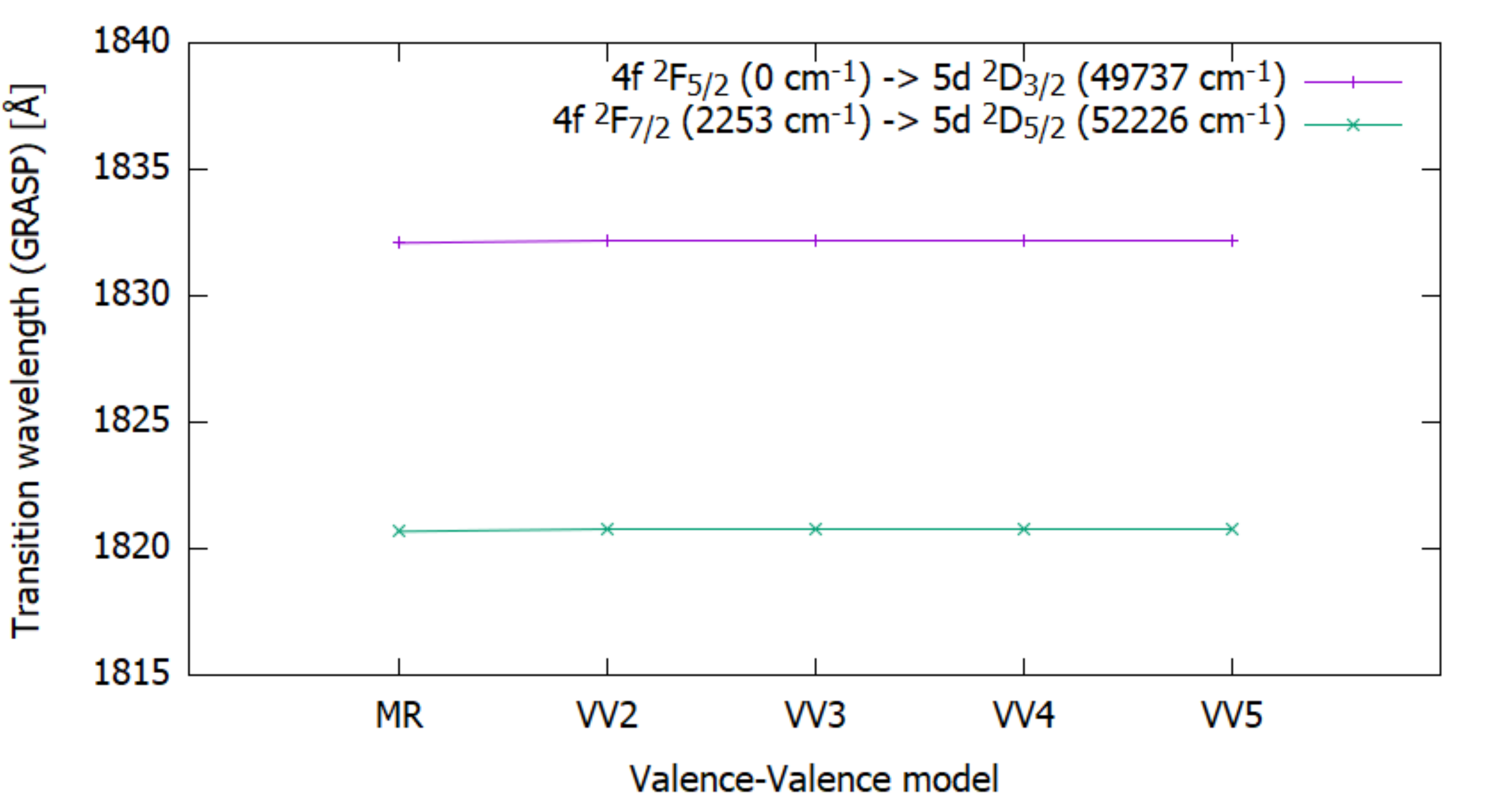}
      \caption{Convergence of transition probabilities calculated in the present work using different valence-valence MCDHF models for two selected Ce IV lines.}
         \label{Fig1}
   \end{figure*}

            \begin{figure*}
   \includegraphics[width=15cm,clip]{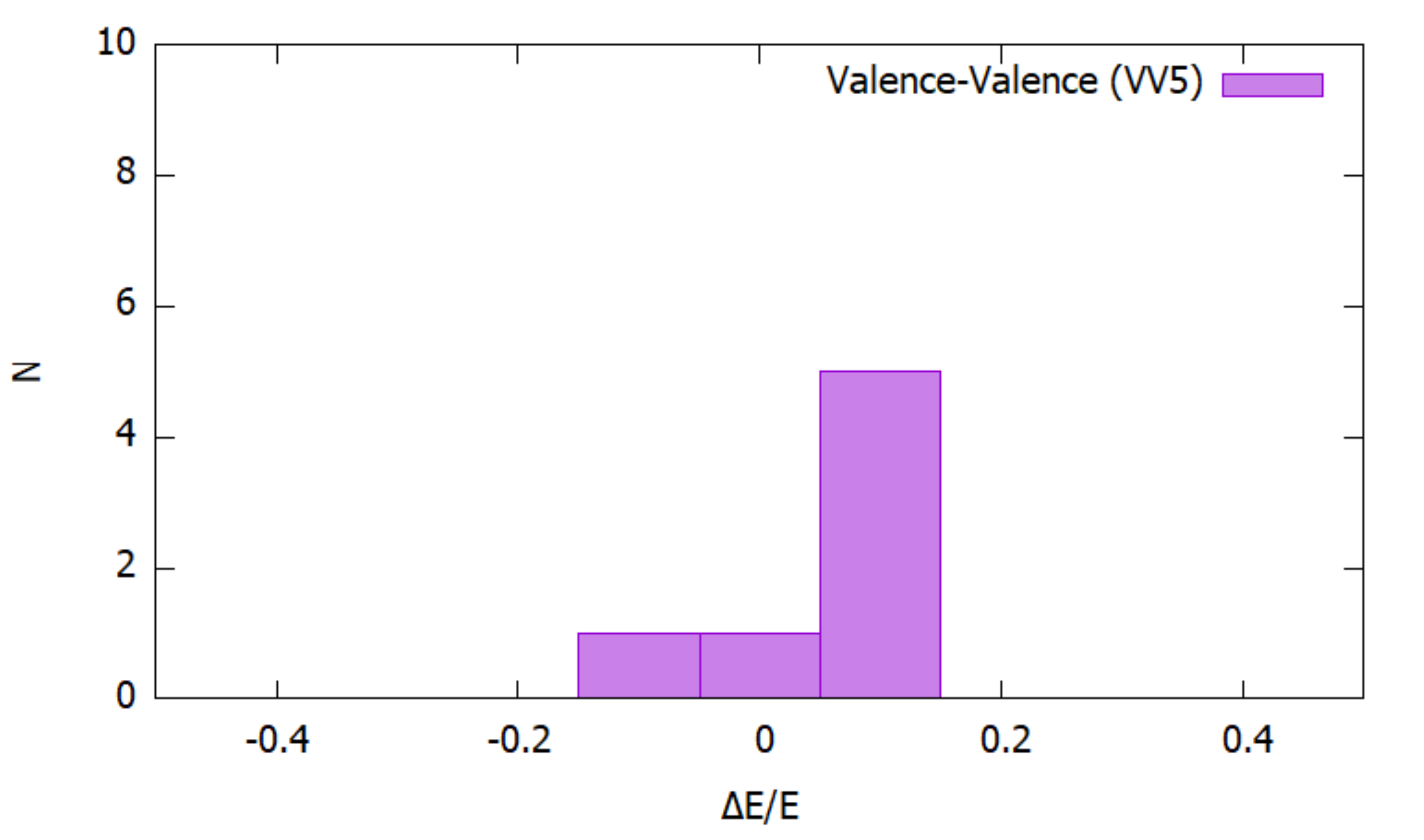}
      \caption{Distribution of energy levels ($N$) according to the mean deviation $\Delta$ E$/$E with the NIST data for Ce IV using our VV5 model.}
         \label{Fig1}
   \end{figure*}

            \begin{figure*}
   \includegraphics[width=15cm,clip]{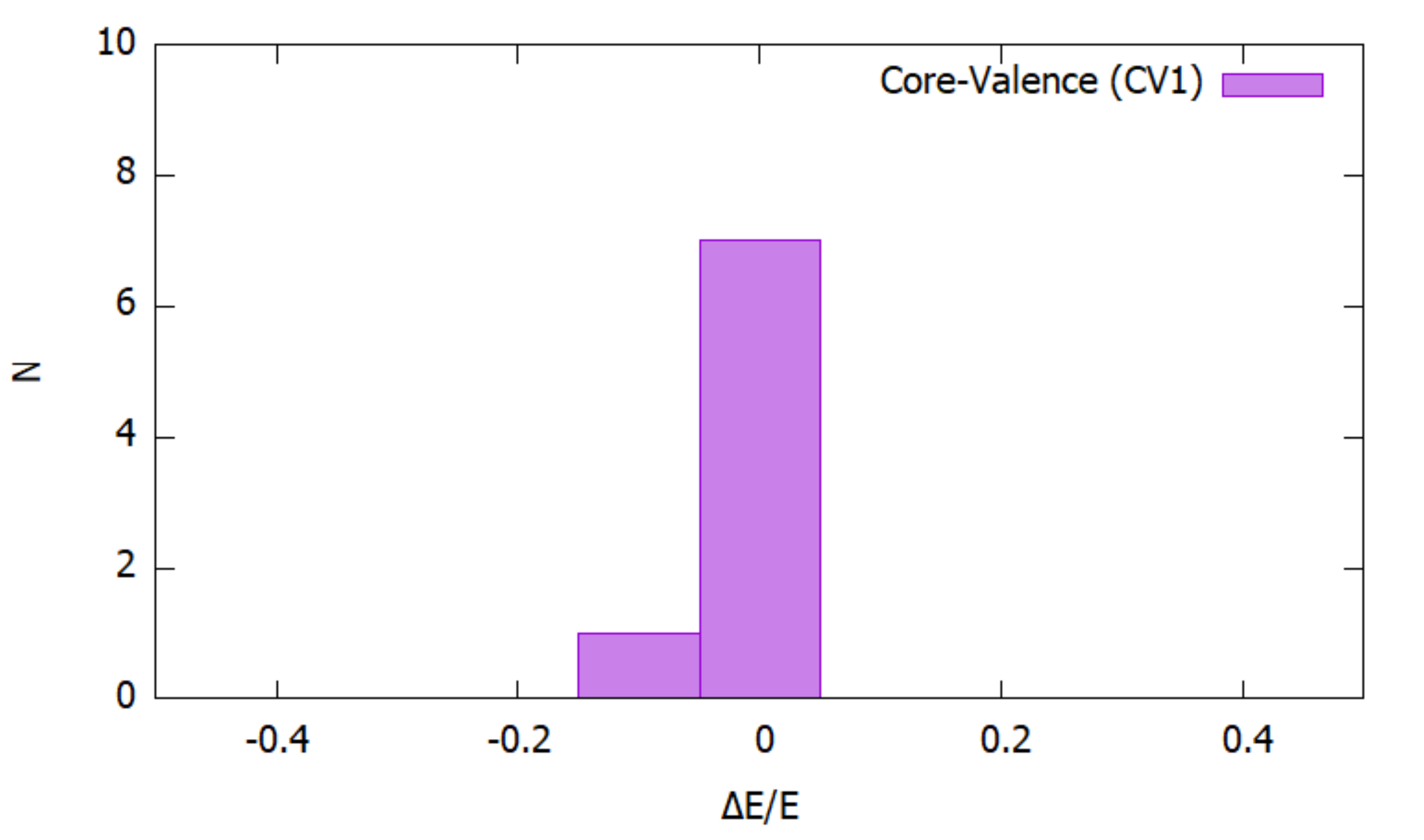}
      \caption{Distribution of energy levels ($N$) according to the mean deviation $\Delta$E$/$E with the NIST data for Ce IV using our CV1 model.}
         \label{Fig1}
   \end{figure*}

              \begin{figure*}
   \includegraphics[width=15cm,clip]{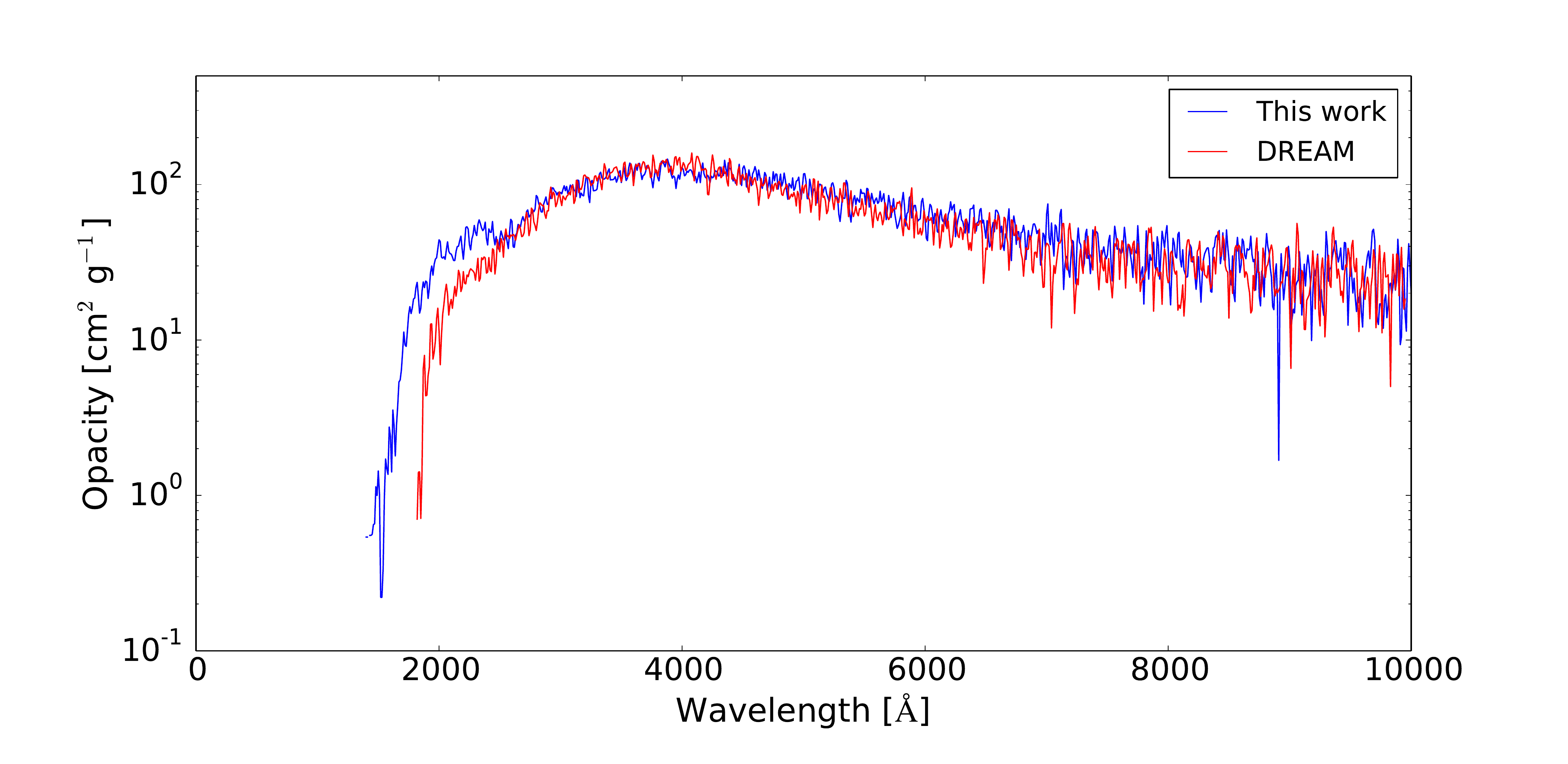}
      \caption{Expansion opacity for Ce II, calculated with $T$ = 5000 K, $\rho$ = 10$^{-13}$ g cm$^{-3}$, $t$ = 1 day and $\Delta$ $\lambda$ = 10 \AA.}
         \label{Fig1}
   \end{figure*}

              \begin{figure*}
   \includegraphics[width=15cm,clip]{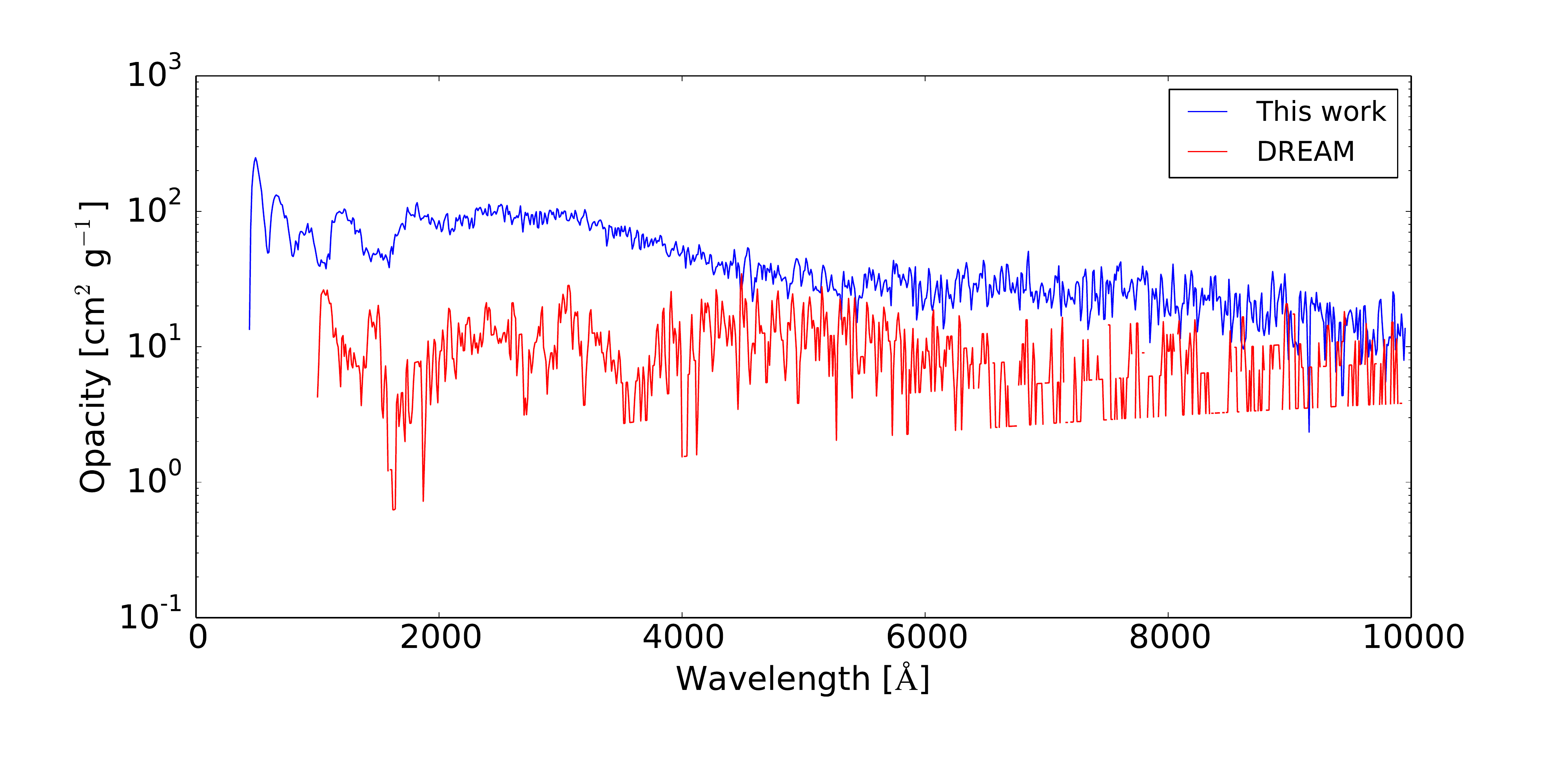}
      \caption{Expansion opacity for Ce III, calculated with $T$ = 10000 K, $\rho$ = 10$^{-13}$ g cm$^{-3}$, $t$ = 1 day and $\Delta$ $\lambda$ = 10 \AA.}
         \label{Fig1}
   \end{figure*}

                 \begin{figure*}
   \includegraphics[width=15cm,clip]{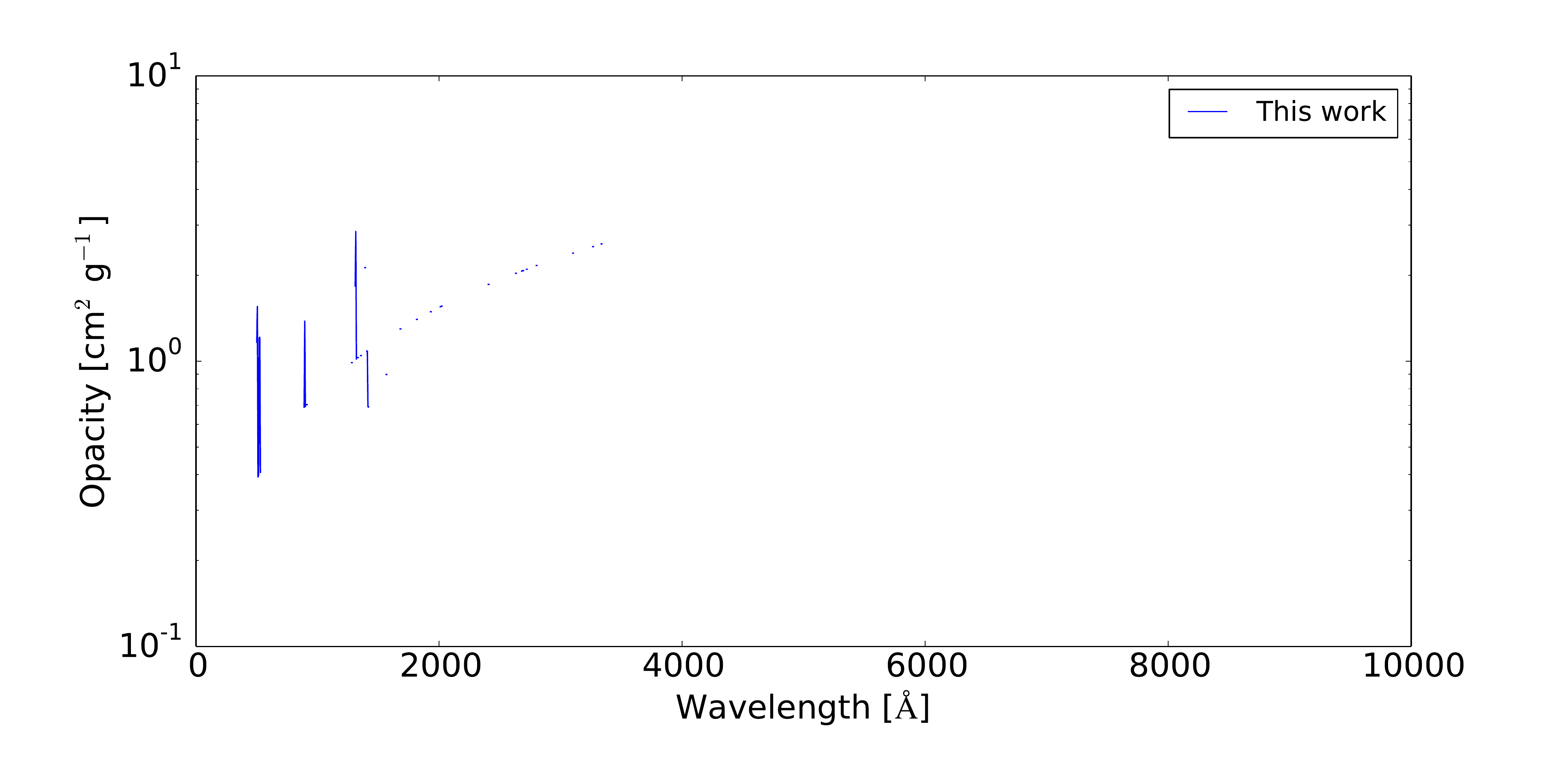}
      \caption{Expansion opacity for Ce IV, calculated with $T$ = 15000 K, $\rho$ = 10$^{-13}$ g cm$^{-3}$, $t$ = 1 day and $\Delta$ $\lambda$ = 10 \AA.}
         \label{Fig1}
   \end{figure*}

                 \begin{figure*}
   \includegraphics[width=15cm,clip]{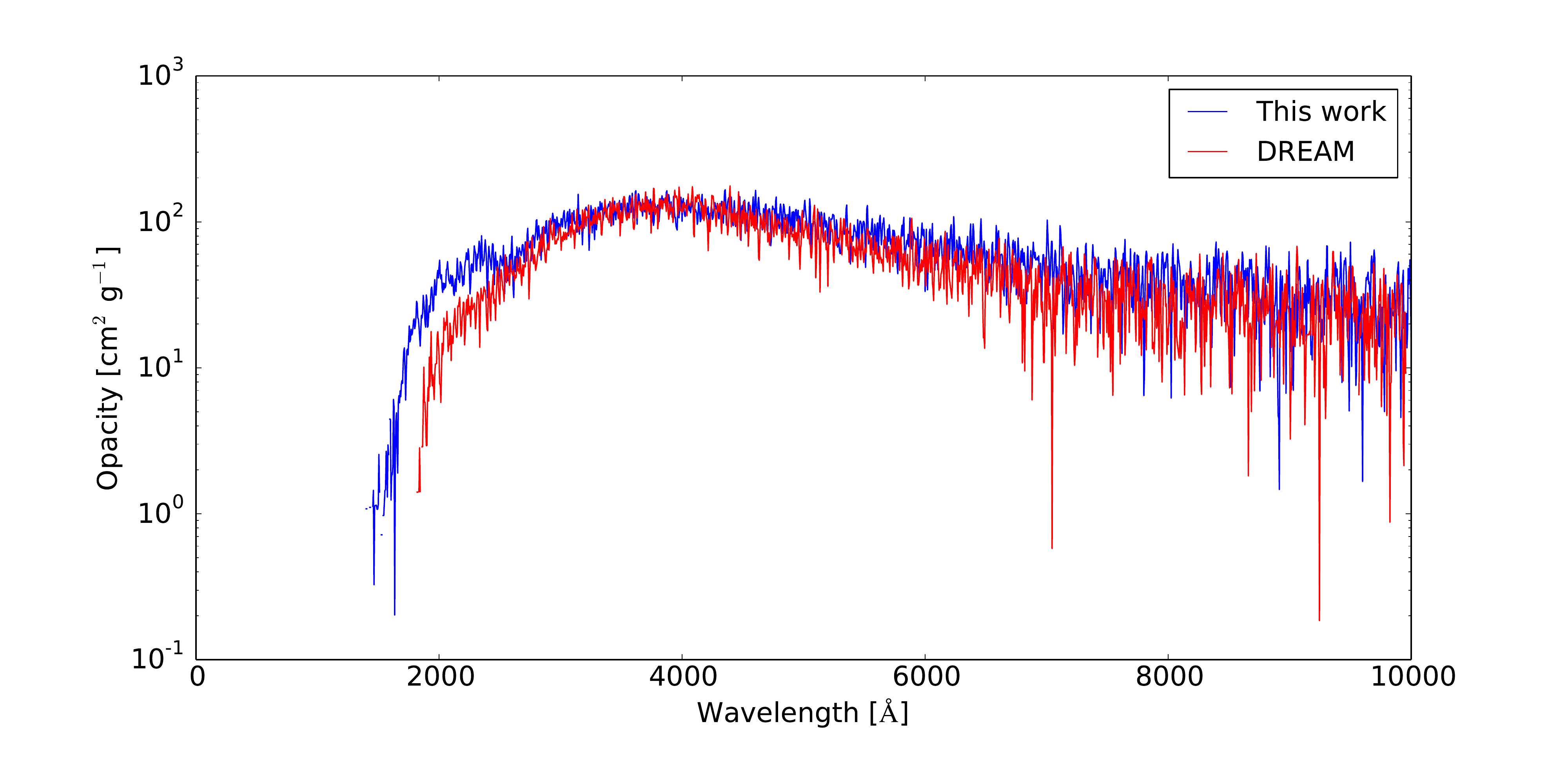}
      \caption{Expansion opacity for Ce II, calculated with $T$ = 5000 K, $\rho$ = 10$^{-13}$ g cm$^{-3}$, $t$ = 1 day and $\Delta$ $\lambda$ = 5 \AA.}
         \label{Fig1}
   \end{figure*}

                 \begin{figure*}
   \includegraphics[width=15cm,clip]{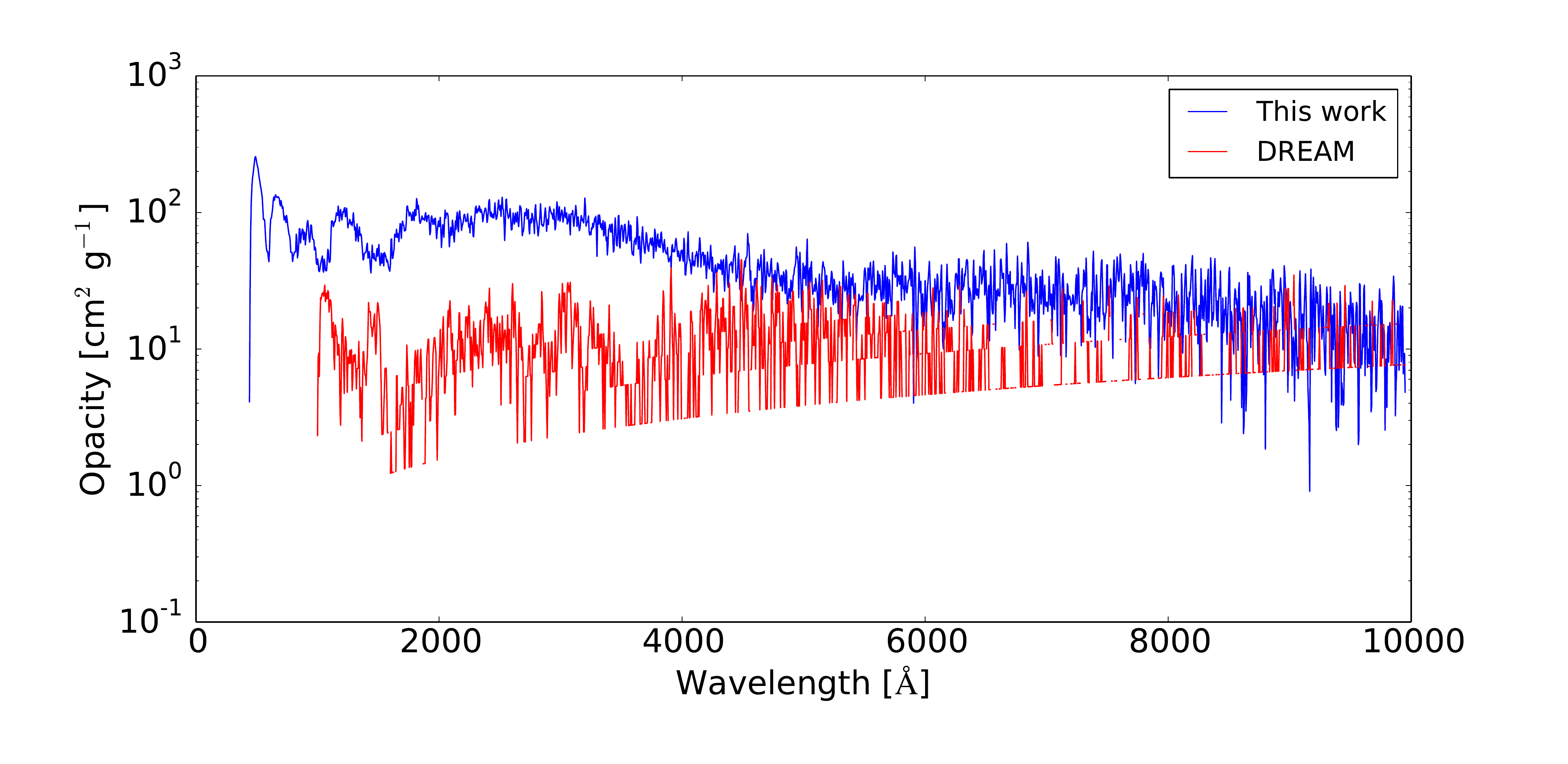}
      \caption{Expansion opacity for Ce III, calculated with $T$ = 10000 K, $\rho$ = 10$^{-13}$ g cm$^{-3}$, $t$ = 1 day and $\Delta$ $\lambda$ = 5 \AA.}
         \label{Fig1}
   \end{figure*}

                    \begin{figure*}
   \includegraphics[width=15cm,clip]{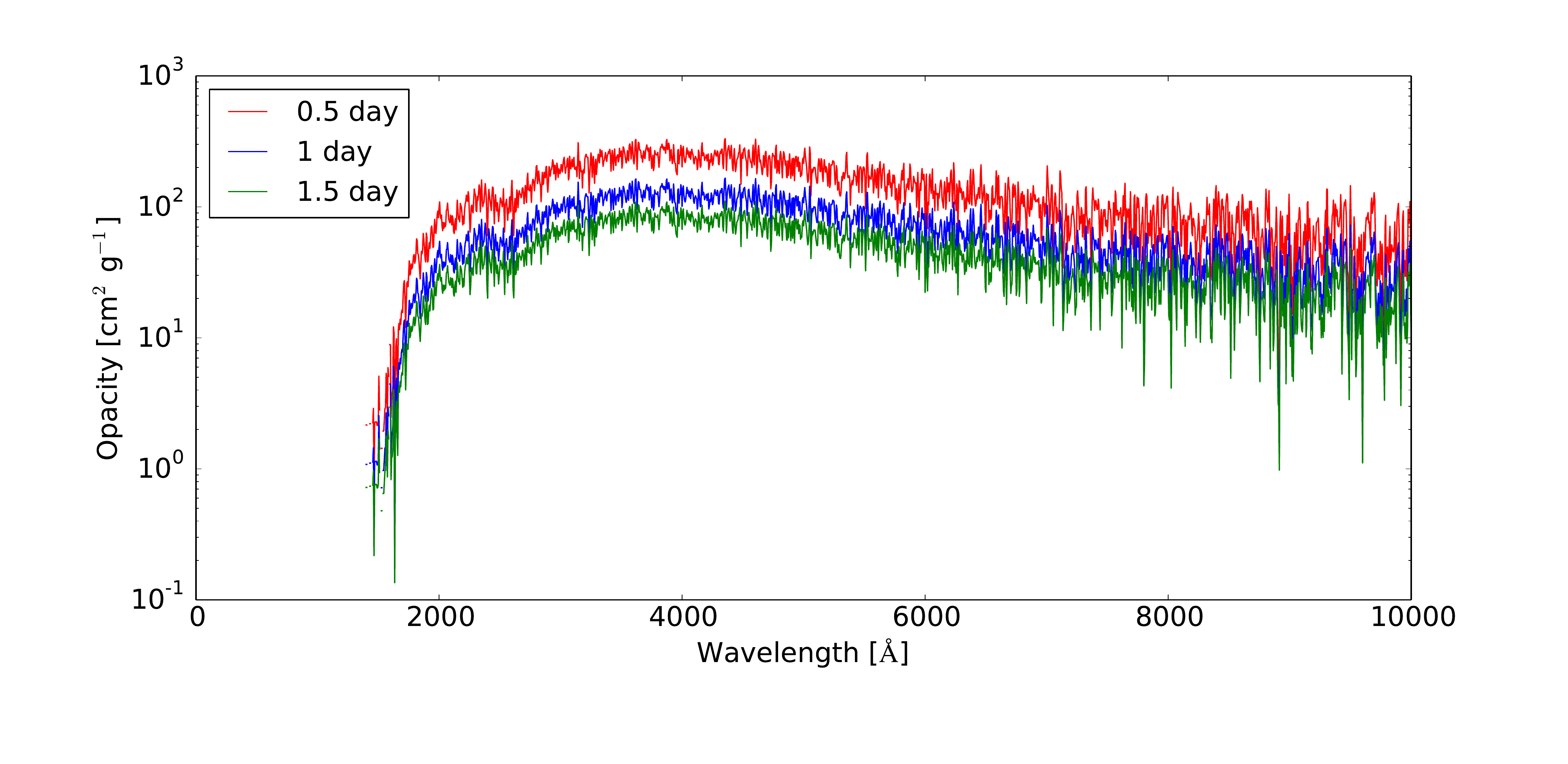}
      \caption{Expansion opacity for Ce II, calculated with $T$ = 10000 K, $\rho$ = 10$^{-13}$ g cm$^{-3}$, $t$ = 0.5 day (red), $t$ = 1 day (blue), $t$ = 1.5 days (green) and $\Delta$ $\lambda$ = 5 \AA.}
         \label{Fig1}
   \end{figure*}

                    \begin{figure*}
   \includegraphics[width=15cm,clip]{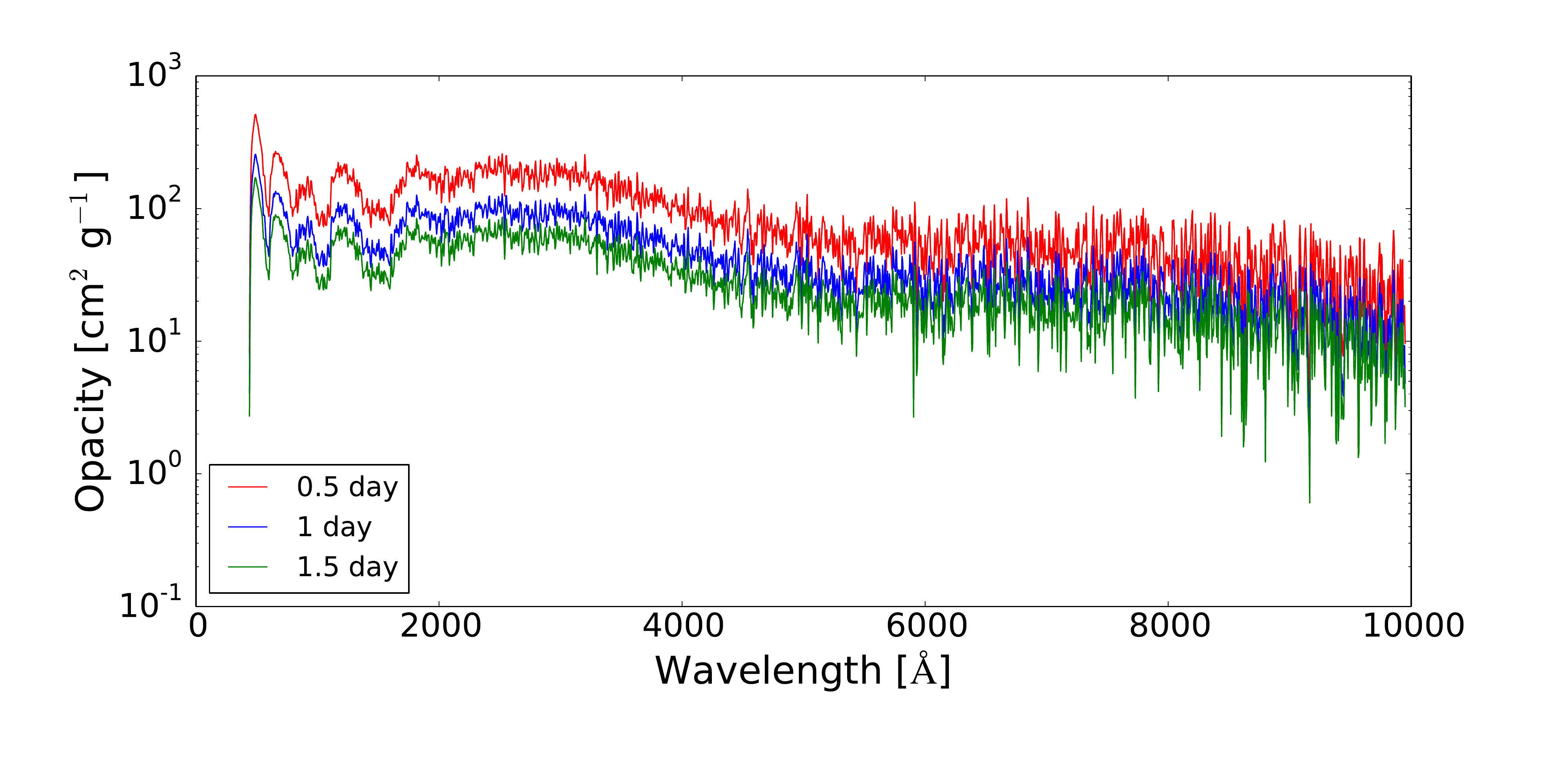}
      \caption{Expansion opacity for Ce III, calculated with $T$ = 10000 K, $\rho$ = 10$^{-13}$ g cm$^{-3}$, $t$ = 0.5 day (red), $t$ = 1 day (blue), $t$ = 1.5 days (green) and $\Delta$ $\lambda$ = 5 \AA.}
         \label{Fig1}
   \end{figure*}

\bsp	
\label{lastpage}


\begin{thebibliography}{99\kern\bibindent}
\makeatletter
\def\@biblabel#1{}
\let\old@bibitem\bibitem
\def\bibitem#1{\old@bibitem{#1}\leavevmode\kern-\bibindent}
\makeatother

\bibitem{}
Abbott B.P., Abbott R., Abbott T.D. {\it et al}, 2017, Phys. Rev. Lett. 119, 161101

\bibitem{}
Bar Shalom A., Klapisch M. and Oreg J., 2001, J. Quant. Spectrosc. Radiat. Transf. 71, 169

\bibitem{}
Cowan R.D., 1981, The Theory of Atomic Structure and Spectra (California University Press,
Berkeley)

\bibitem{}
Eastman R.G. and Pinto P.A., 1993, Astrophys. J. 412, 731

\bibitem{}
Froese Fischer C. and Godefroid M.R., 2019, Phys. Rev. A 99, 032511

\bibitem{}
Froese Fischer C., Godefroid M., Brage T., J\"onsson P., Gaigalas~G., 2016, J. Phys. B: At. Mol. Opt. Phys. 49, 182004

\bibitem{}
Froese Fischer C., Gaigalas G., J\"onsson, and J. Biero\'n, 2019, Comput. Phys. Commun. 237, 184

\bibitem{}
Gaigalas G., Kato D., Rynkun P., Rad{\v{z}}i{\={u}}t{\.{e}} L. and Tanaka M., 2019, Astrophys. J. Suppl. 240, 29

\bibitem{}
Gaigalas G., Rynkun P., Rad{\v{z}}i{\={u}}t{\.{e}} L., Kato D., Tanaka~M. and J\"onsson P., 2020, Astrophys. J. Suppl. 248, 13

\bibitem{}
Grant I.P., 2007, Relativistic Quantum Theory of Atoms and Molecules (Springer)

\bibitem{}
J\"onsson P., Gaigalas G., Biero\'n J., Froese Fischer C., Grant, I., 2013, Comput. Phys. Commun. 184, 2197

\bibitem{}
Karp A.H., Lasher G., Chan K.L. and Salpeter E.E., 1977, Astrophys. J. 214, 161

\bibitem{}
Kasen D., Metzger B., Barnes J. Quataert E. and Ramirez-Ruiz E., 2017, Nature 551, 80

\bibitem{}
Kasen D., Thomas R.C. and Nugent P., 2006, Astrophys. J. 651, 366

\bibitem{}
Kramida A., Ralchenko Yu., Reader J. and NIST ASD Team, 2020, NIST Atomic Spectra Database (ver.5.7.1.), Available online at https://physics.nist.gov/asd (accessed in August 2020)

\bibitem{}
Quinet P. and Palmeri P., 2020, Atoms 8, 18

\bibitem{}
Quinet P., Palmeri P., Bi\'emont E., Li Z.S., Zhang Z.G. \& Svanberg S., 2002, J. Alloys Comp., 344, 255

\bibitem{}
Quinet P., Palmeri P., Bi\'emont E., McCurdy M.M., Rieger G., Pinnington E.H.,
Wickliffe M.E., Lawler J.E., 1999, Mon. Not. R. Astron. Soc., 307, 934

\bibitem{}
Rad{\v{z}}i{\={u}}t{\.{e}} L., Gaigalas G., Kato D., Rynkun P. and Tanaka M., 2020, Astrophys. J. Suppl. 248, 17

\bibitem{}
Sobolev V.V., 1960, Moving Envelopes of Stars (Harvard University Press)

\bibitem{}
Tanaka M., Kato D., Gaigalas G. $et~al.$, 2018, Astrophys. J.  852, 109

\bibitem{}Tanaka M., Kato D., Gaigalas G. and Kawaguchi K., 2020, Mon. Not. R. Astron. Soc. 496, 1369




\end{thebibliography}
\end{document}